\documentclass[a4paper,12pt]{article}
\usepackage[dvips]{graphics}
\usepackage{amssymb}
\usepackage{amsmath}
\usepackage{color}
\usepackage{pstricks}
\usepackage{graphicx}
\usepackage{dcolumn} 
\usepackage{multirow, booktabs}

\usepackage{cite}
\newcommand{\beq}{\begin{equation}}
\newcommand{\eeq}{\end{equation}}
\newcommand{\bea}{\begin{eqnarray}}
\newcommand{\eea}{\end{eqnarray}}
\newcommand{\lsp}{\tilde\chi^0_1}
\newcommand{\mlsp}{m_{\tilde\chi^0_1}}

\newcommand{\grdec}{\widetilde{G} \to \lsp \, X \, Y }
\newcommand{\grav}{\widetilde{G}  }

\newcommand\iso[2]{\mbox{${}^{#2}${\rm #1}}}
\def\he#1{\iso{He}{#1}}

\def\li#1{\iso{Li}{#1}}
\def\b1#1{\iso{B}{1#1}}

\makeatletter

\def\slash{\@ifnextchar[{\fmsl@sh}{\fmsl@sh[0mu]}}
\def\fmsl@sh[#1]#2{%
  \mathchoice
    {\@fmsl@sh\displaystyle{#1}{#2}}%
    {\@fmsl@sh\textstyle{#1}{#2}}%
    {\@fmsl@sh\scriptstyle{#1}{#2}}%
    {\@fmsl@sh\scriptscriptstyle{#1}{#2}}}
\def\@fmsl@sh#1#2#3{\m@th\ooalign{$\hfil#1\mkern#2/\hfil$\crcr$#1#3$}}
\makeatother

\newcommand{\fourvector}[4]{\left(\begin{array}{c} #1\\ #2 \\ #3 \\ #4 \end{array}\right)}
\newcommand{\eightvector}[8]{\left(\begin{array}{c} #1\\ #2 \\ #3 \\ #4 \\ #5\\ #6 \\ #7 \\ #8\end{array}\right)}

\def\a{\alpha}
\def\b{\beta}
\def\c{\chi}

\def\e{\epsilon}

\def\g{\gamma}

\def\l{\lambda}
\def\m{\mu}
\def\n{\nu}

\def\r{\rho}

\newcommand{\mgrav}{m_{\widetilde{G}}}

\newcommand{\mplanck}{\ensuremath{M_{\text{P}}}}

\newcommand{\cf}[2]{\c_{#1}^{#2}}   
\newcommand{\cfb}[2]{\overline{\c}_{#1}^{#2}}   
\newcommand{\gen}[3]{T_{#1,\, #2 #3\,}}  

\newcommand{\gbalpha}[2]{{A}^{(#1)\,#2}}   

\def\slash#1{\rlap{\hbox{$\mskip 1 mu /$}}#1}   

\definecolor{darkgreen}{rgb}{0,.5,0}

\begin{document} 
\begin{titlepage} 
 
\begin{flushright}
HEPHY-PUB 924/13\\  
 \end{flushright} 
\begin{center} 
\vspace*{1.5cm} 
 
{\Large{\textbf {Three-body  gravitino decays in the MSSM}}}\\ 
\vspace*{10mm} 
 
{\bf  Helmut~Eberl}$^1$ and  {\bf  Vassilis~C.~Spanos}$^2$ \\ 

\vspace{.7cm} 

$^1${{\it Institut f\"ur Hochenergiephysik der \"Osterreichischen Akademie
der Wissenschaften,}\\
{\it A--1050 Vienna, Austria}

 $^2${\it Institute of Nuclear and Particle Physics, NCSR ``Demokritos'', GR-15310 Athens, Greece}} \\

\end{center} 
\vspace{2.cm}
 
\begin{abstract} 
We present results from a new calculation of two- and three-body decays of the gravitino including all possible 
Feynman graphs. The work is done in the $R$-parity conserving Minimal Supersymmetric Standard Model,
assuming that the gravitino is
unstable. We calculate the full width for the three-body decay process $\grdec$, where $\lsp$ is the
lightest neutralino that plays the role of the lightest supersymmetric particle, and $X,\, Y$ are standard model 
particles. For calculating the squared amplitudes we use the packages {\tt FeynArts} and {\tt FormCalc}, 
with appropriate extensions to accommodate the gravitino with spin~$3/2$.
Our code treats automatically the intermediate exchanged particles that are on their mass shell,
using the narrow width approximation. This method considerably simplifies  
the complexity of the evaluation of the total gravitino width and stabilizes it numerically. 
At various points of the supersymmetric parameter space, we compare the full two-body and the 
three-body results with approximations used in the past.
In addition, we discuss the relative sizes of the full two-body and the three-body gravitino decay widths.  
Our results are obtained using the package {\tt GravitinoPack}, that will be publicly available soon,
including various handy features.
\end{abstract}

\end{titlepage} 
\baselineskip=18pt
\section{Introduction}
\label{sec:intro}
\label{intro}
The gravitino, the supersymmetric partner of the graviton,
is  part of the particle spectrum of extensions of the Standard Model (SM)
that incorporate the local version of  supersymmetry, the supergravity.
Depending on the mass hierarchy of these models, 
in the case of $R$-parity conservation,  
the gravitino can be either the stable 
lightest supersymmetric particle  (LSP)  that  plays the role
of the dark matter particle, or it is heavier than the LSP and thus unstable. 
In the latter case, it is important to calculate precisely the width of the 
gravitino decays to the lightest neutralino ($\lsp$), that naturally is the LSP in this case,  and other SM particles.
The precise value of the gravitino lifetime is probably the most  
important parameter  that affects the  primordial Big-Bang
Nucleosynthesis (BBN) constraints, assuming that it decays during or after the BBN time. 
Indeed, the abundances of the light nuclei, like D, \he4, \he3 and \li7  produced 
during the BBN epoch,  provide very stringent constraints on
the decays of unstable massive particles during the early
Universe~\cite{Lindley:1984bg, Ellis:1984er, Lindley:1986wt, Scherrer:1987rr, Reno, 
Dimopoulos:1988ue, ellis, Kawasaki:1994af,kawmoroi, kkm,  kohri, cefo, jed, kkm2,Cyburt:2009pg,MS1}.
 
The   astrophysical  measurements  of the abundances of various light elements  agree
well with those predicted by the homogeneous BBN calculations. Thereby, it is assumed that the
baryon-to-photon ratio $\eta$ required  for these
calculations~\cite{cfo1} agrees very well with that
inferred~\cite{cfo2} from observations of the power spectrum of
fluctuations in the cosmic microwave background. On the other hand, the value of $\eta$
that they provide is now quite precise, reducing the main uncertainty  in the  BBN calculation.
The decays of massive unstable particles, like the gravitino,  could have destroyed this
concordance because the electromagnetic and/or hadronic showers they
produced might have either destroyed or created the light nuclear species.
In this calculation, probably  the most important parameter is the lifetime of the unstable gravitino,
because it dictates the cosmic time when these injections will occur~\cite{Cyburt:2009pg}.

In the previous studies  the dominant two-body decays of the gravitino have been considered,
that is the processes $\grav \to \widetilde{X}\, Y$, where $\widetilde{X}$ is a sparticle, and $Y$ a SM particle. These two-body
decays dominate the total gravitino width, and in particular the channel $\grav \to  \lsp \, \gamma $,  which  is kinematically open 
in  the whole region of the neutralino LSP model,   $\mgrav > \mlsp$. This 
channel is  also important because  it provides the bulk of the electromagnetic  energy produced
by the gravitino decays. On the other hand, the decay $\grav \to  \lsp \, Z $ is the dominant channel that produces 
hadronic products. 
Since, these decays are of ``gravitational nature",   the gravitino 
width is proportional to $M_P^{-2}$, where $M_P=1/\sqrt{8 \pi G_N}$ is the reduced Planck mass, and its lifetime is 
of the order of few seconds or more. Afterwards, the produced sparticle  $\widetilde{X}$ decays almost instantaneously in
comparison to the gravitational decay, like $ \widetilde{X} \to \lsp \, X$, where $X$ is a SM particle. 
So, assuming that   the dominant two-body channel is kinematically possible,  $\mgrav > m_{\widetilde{X}}  + m_Y$, the 
gravitino decays  to the LSP $\lsp$ as  
\beq
\grav \xrightarrow{\, \, grav\, \, }  \widetilde{X}\, Y \xrightarrow{\, \, EW\, \, }   \lsp \, X \,Y \, ,
\label{eq:2B}
\eeq
 where, as it is indicated,  the first decay is of gravitational and the second one of electroweak nature.  
If  the primary two-body gravitino decay channel is not open, $\mgrav < m_{\widetilde{X}}  + m_Y$, the three-body
decay 
\beq
\grav  \xrightarrow{\, \, grav\, \, }   \lsp \, X \,Y 
\label{eq:3B}
\eeq
is still possible, if $\mgrav > \mlsp+ m_X +m_Y $,   through   off-shell sparticles exchange. 
Therefore, it is important to include the three-body channels, like $\grav \to \lsp \, q \, \bar{q}$, which
inject hadrons in the cosmic plasma even below the threshold for the on-shell $Z$-boson  production.
This  channel, together with   the $\grav \to \lsp \, W^+ \, W^-$, usually have been taken into account in the previous studies. 
In particular,  the channel $ \lsp \, W^+ \, W^-$ and  its high energy behavior  has been already studied in detail \cite{Cyburt:2009pg,Luo:2010he},  
due to the tricky cancellation of the unitarity violating terms that are associated with the longitudinal components of the $W$-bosons. 
 
The purpose of this paper is to present the complete  calculation of the gravitino  two-body decays,     {\em as well as  all the possible  three-body }
decay channels $\grav \to \tilde \chi^0_1 \, X \, Y$.
This calculation provides us with a comprehensive picture of the gravitino decays above and {\em below } 
the kinematical thresholds of the two-body decays. 
For the computation of the gravitino decay amplitudes we use the {\tt  FeynArts} ({\tt FA}) and {\tt FormCalc} ({\tt FC}) packages~\cite{FAref,  FAFCref , LTref }. 
In order to be able to generate the amplitudes with {\tt FA} we have extended its generic structures to interactions with spin-3/2 particles and built a model file with all possible gravitino interactions with the particles of the Minimal Supersymmetric Standard Model (MSSM). 
{\tt FC} has been extended so that it automatically generates a {\sc Fortran}
 code for the numerical calculation
of the squared amplitudes. As there are many gamma matrices involved we have done this within {\tt FC} by using the Weyl-van-der-Waerden 
formalism~\cite{WvdW}, as implemented into {\tt FC} from~\cite{Dittmaier:1998nn}.

Moreover, the code treats automatically the intermediate exchanged particles  that are on their mass shell as unstable, 
using the narrow width approximation. 
By doing so, one avoids the double counting among  the two- and three-body channels, but also the
{\sc cpu}-time consuming phase space integrations over the unstable exchanged particles in the propagators. 
It also leads to a considerable 
simplification of the numerical evaluation of the  total three-body gravitino width. 
 
Our computer code will become publicly available soon as part of {\tt GravitinoPack}~\cite{GravPackpaper}.
This package supports the SUSY Les Houches Accord 2 format~\cite{SLHAref, SLHA2ref} and works within the general flavour conserving MSSM  with possibly complex input parameters.
It incorporates handy {\sc Fortran}  and  {\sc Mathematica} routines that evaluate two- and three-body gravitino decay widths.
The code returns all partial decay widths (or branching ratios) for
the individual decay channels.

As basis for our numerical study  we use various benchmark points from supersymmetric models with different characteristics. 
In  particular, we choose  points based 
on the phenomenological MSSM (pMSSM),   the 
Constrained MSSM (CMSSM)  and the  Non-Universal Higgs Model (NUHM). 
These points satisfy all the recent high energy constraints from the LHC experiments on the superpartner lower mass bounds,
the Higgs boson mass $\sim 126$~GeV, as well as the LHCb measurement on the rare decay $B_s \to \mu^+ \mu^-$ that
constrains supersymmetric models at the large $\tan\beta$ regime. In addition, these points conform with the cosmological constraints on
the amount of the neutralino relic density, measured by the Planck  and the WMAP experiments.

Numerically we have found that the calculation of the full three-body decay amplitudes is important  
in the computation of 
the gravitino width for any gravitino mass, above and below the mass thresholds for the dominant two-body decay channels.
Furthermore, we have compared the total gravitino decay width, up to the three-body level, 
with a previously used approximation. We have found  that the full result can differ from the approximation by more 
than a factor of 2, especially  for $\mgrav  \gtrsim $ 2 TeV.
This affects considerably the gravitino lifetime, which is an important parameter in
constraining models with unstable gravitinos, using the BBN predictions.
Interestingly enough, we have found that both goldstino ($\pm {1 \over 2}$) and the
pure gravitino ($\pm {3 \over 2}$) spin states  contribute comparably, even for large gravitino masses,  up to $6$ TeV or  larger.

The  paper  is organized as: In section~\ref{sec:calculation}
the details of our calculation are explained. In particular, we analyse  the gravitino interactions in the MSSM and
we demonstrate, using specific examples, our implementation in the {\tt FA}/{\tt FC} code. Moreover,  the calculation of the 
gravitino two- and three-body decay width  is discussed.
In section~\ref{sec:numerics} we present the numerical results, using as basis various benchmark points from different supersymmetric models.
Finally in section~\ref{sec:conclusions}, we summarize our findings and give a perspective of our future work in this direction. 

\section{The calculation}
\label{sec:calculation}

\subsection{Gravitino interactions within the MSSM}
\label{subsec:inter}
Our starting point is the Lagrangian that describes
the gravitino interactions in the MSSM ~ \cite{PradlerPhD},
\begin{align}
  \label{eq:gravitino interaction lagrangian}
  \mathcal{L}^{(\a) }_{{\widetilde G},\,\text{int}} &= -\frac{i}{\sqrt{2}\mplanck} \left[
  \mathcal D^{(\a) }_{\m} \phi^{*i} \overline{{\widetilde G}}_\n \g^\m \g^\n \cf Li
  -\mathcal D^{(\a) }_{\m} \phi^{i} \cfb Li \g^\n \g^\m  {\widetilde G}_\n
\right]
 \nonumber \\
  & \hspace*{2cm} -  \frac{i}{8\mplanck}\overline{\widetilde G}_\m [\g^\n,\g^\r] \g^\m
  {\l}^{(\a)\, a} {F}_{\n\r}^{(\a)\, a}\; ,
\end{align}
with the covariant derivative given by
\begin{equation}
   \label{eq:covariant derivative}
    \mathcal{D}^{(\a)}_\mu \phi^i =
    \partial_\mu \phi^i + i {g}_\a \gbalpha {\a}{a}_\m \gen aij^{(\a)} \phi^j\; ,
\end{equation}
and the field strength tensor ${F}_{\m\n}^{(\a)\,a}$ reads
\begin{align}
  \label{eq:field strength tensor}
  {F}_{\m\n}^{(\a)\,a} & = \partial_\m {A}_\n^{(\a)\,a} -
  \partial_\n {A}_\m^{(\a)\,a} -{g}_\a f^{(\a)\,abc} {A}_\m^{(\a)\,b} {A}_\n^{(\a)\,c}\; .
\end{align}
The index $\a$ corresponds to the three groups U(1)$_Y$, SU(2)$_L$, and SU(3)$_c$ with
$a$ up to  $1, 3, 8$, respectively, and $i = 1, 2, 3$. $\phi^i, \chi^j, A^a$, and $\lambda^a$ are the scalar, spin~1/2, gauge and
gaugino fields of the MSSM, respectively, as given in the Tables 1 and 2 in \cite{PradlerPhD}.

We write the four polarisation states of the gravitino field in the momentum space in terms 
of spin-1 and spin-1/2 components as
\begin{eqnarray}
\psi_\mu({\bf k}, {3 \over 2}) & = & u({\bf k} , {1 \over 2}) \,  \e_\mu({\bf k}, 1)\, , \nonumber\\
\psi_\mu({\bf k}, {1 \over 2}) & = & \sqrt{1 \over 3}  \,  u({\bf k}, -{1 \over 2})  \,   \e_\mu({\bf k}, 1)+
                                     \sqrt{2 \over 3}  \,  u({\bf k}, {1 \over 2})  \,   \e_\mu({\bf k}, 0)\,  \nonumber\\
\psi_\mu({\bf k}, -{1 \over 2}) & = & \sqrt{1 \over 3}  \,  u({\bf k}, {1 \over 2})  \,   \e_\mu({\bf k}, -1)+
                                     \sqrt{2 \over 3}  \,  u({\bf k}, -{1 \over 2})  \,  \e_\mu({\bf k}, 0)\, \nonumber\\
\psi_\mu({\bf k}, -{3 \over 2}) & = & u({\bf k}, -{1 \over 2})  \,  \e_\mu({\bf k}, -1)\, ,
\end{eqnarray}
with $\bf k$ being the spatial part of the four momentum  $k^\mu = (k^0, |{\bf k|} {\bf e})$, and 
${\bf e}$ the unit vector in the flight direction of the gravitino.

From the field equation for the spin-3/2 particle, the so called
Rarita-Schwinger equation \cite{Rarita-Schwingerref}, we get three equations in momentum space,
which we have to fulfill by the right choice of the spin-1 and spin-1/2
field components \cite{higherspinref},
\begin{eqnarray}
\g^\m \psi_\mu({\bf k}, \l) & = & 0\,, \label{Rarita-Schwinger1}\\
k^\m \psi_\mu({\bf k}, \l) & = & 0\,, \label{Rarita-Schwinger2}\\
(\slash{k} - m_{\widetilde G}) \psi_\mu({\bf k}, \l) & = & 0 \label{Rarita-Schwinger3}\, , \
\end{eqnarray}
where $m_{\widetilde G}$ is the mass of the gravitino. We have checked that our implementation of 
$\psi_\mu$ into {\tt FC} indeed fulfills these three equations.

\begin{figure}[t!]
\begin{center}
\begin{tabular}{rl}
\resizebox{6.3cm}{!}{\includegraphics{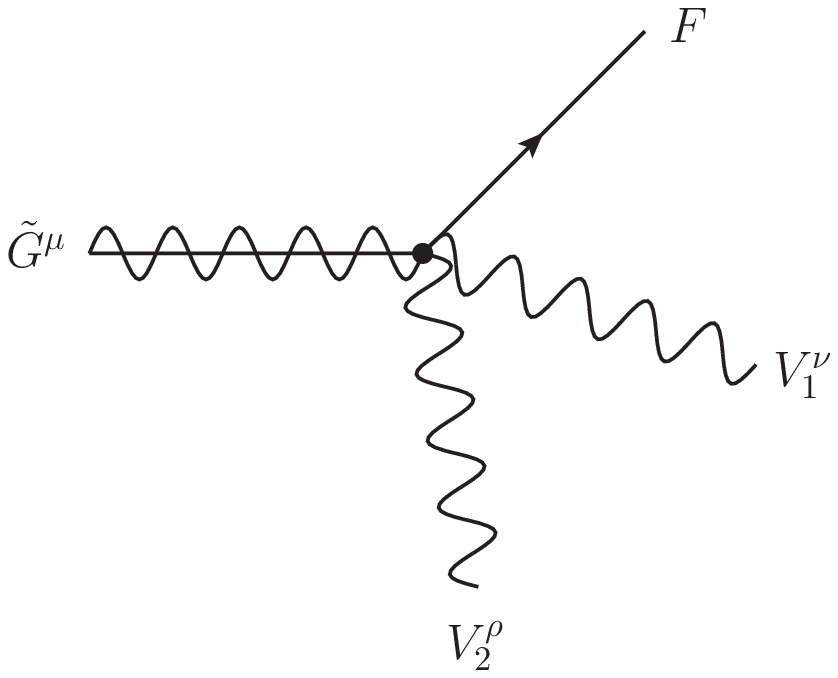}}
&
\raisebox{3cm}{
${\rm struc1:}  \fourvector{\g^\mu \left[\g^\nu,\, \g^\rho \right] P_L}{\g^\mu \left[\g^\nu,\, \g^\rho \right] P_R}
                           {\left[\g^\nu,\, \g^\rho \right] \g^\mu P_L}{\left[\g^\nu,\, \g^\rho \right] \g^\mu P_R}
. C[F|\bar F, \widetilde G^\mu, V_1^\nu, V_2^\rho]
$
}\\
\resizebox{6.3cm}{!}{\includegraphics{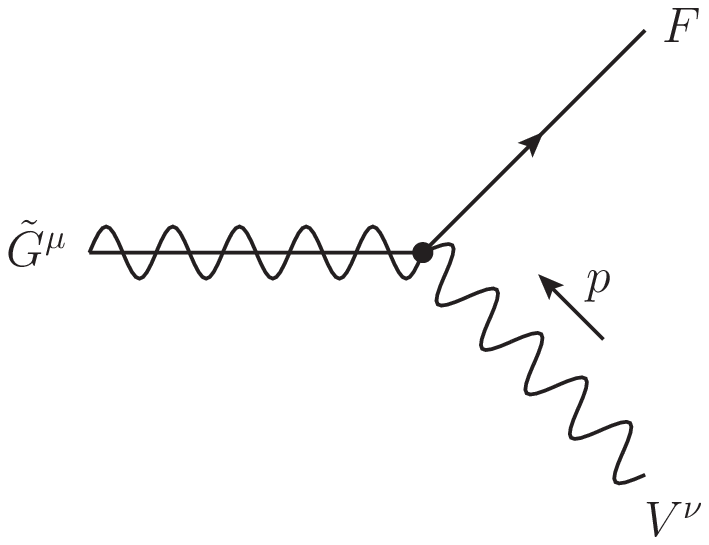}}
&
\raisebox{2.5cm}{
${\rm struc2:}  \eightvector{\g^\mu \g^\nu P_L}{\g^\mu \g^\nu P_R}
                           {\g^\mu \left[\g^\nu,\, \slash{p}\,\right] P_L}
                           {\g^\mu \left[\g^\nu,\, \slash{p}\,\right] P_R}
                           {\g^\nu \g^\mu P_L}{ \g^\nu \g^\mu P_R}
                           {\left[\g^\nu,\, \slash{p}\,\right] \g^\mu P_L}
                           {\left[ \g^\nu,\, \slash{p}\,\right] \g^\mu P_R}
. C[F|\bar F, \widetilde G^\mu, V^\nu]
$}\\
\resizebox{6.3cm}{!}{\includegraphics{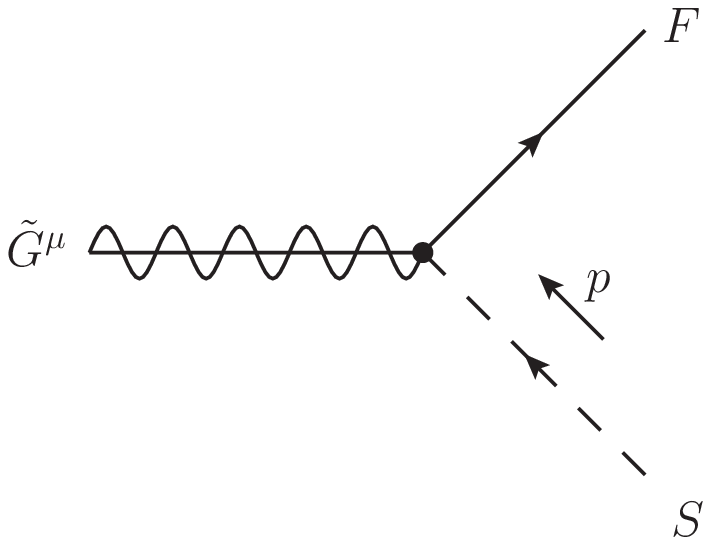}}
&
\raisebox{2.5cm}{
${\rm struc3:} \fourvector{\g^\mu \slash{p}\, P_L}{\g^\mu \slash{p}\, P_R}
                          {\slash{p} \g^\mu \, P_L}{\slash{p} \g^\mu \, P_R}
. C[F|\bar F, \widetilde G^\mu, S]
$}\\
\resizebox{6.3cm}{!}{\includegraphics{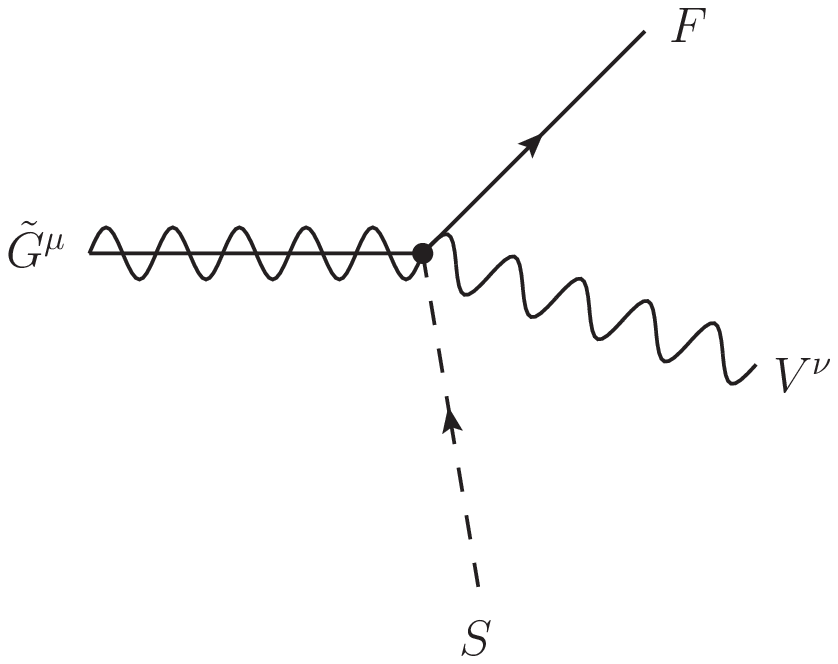}}
&
\raisebox{3cm}{
${\rm struc4:} \fourvector{\g^\mu \g^\nu P_L}{\g^\mu \g^\nu P_R}
                          {\g^\nu \g^\mu P_L}{\g^\nu \g^\mu P_R}
. C[F|\bar F, \widetilde G^\mu, V^\nu, S]
$}
\end{tabular}
\end{center}
\caption{Possible structures of gravitino interactions with MSSM particles, detailed explanation
is given in the text. As all momenta are defined incoming in {\tt FA}, $\partial_\mu \to -i p_\mu$. 
\label{fig:FAstructures1}}
\end{figure}
\begin{figure}[t!]
\begin{center}
\begin{tabular}{rl}
\resizebox{7cm}{!}{\includegraphics{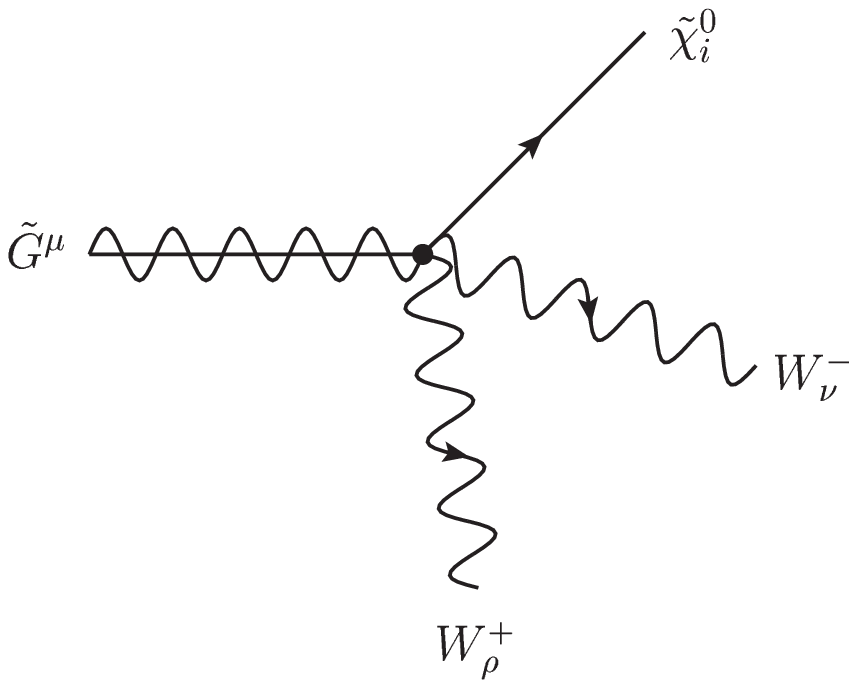}}
&
\raisebox{3cm}{\hspace*{-1cm}
$i {g \over 4 \mplanck}  \g^\m [\g^\n,\g^\r]
 (Z_{i,2} P_L + Z^*_{i,2} P_R) $}
\end{tabular}
\end{center}
\caption{Feynman rule for ${\widetilde G} \to \tilde \chi^0_i \, W^+ \, W^-$ four-particle interaction that
corresponds to the Lagrangian in  Eq.~(\ref{vertexNGrWWhc}).
\label{fig:vertexNeuGrWW}}
\end{figure}

All possible structures for the gravitino interactions with the  MSSM particles in Eq.~(\ref{eq:gravitino interaction lagrangian}), 
are depicted in Figure~\ref{fig:FAstructures1}.
We have extended the {\tt FA} generic Lorentz file with these structures. The 78 couplings given by the 
coefficient vectors $C[\ldots]$, have been appended to the {\tt FA} MSSM model file. 
Further technical  details will be provided in \cite{GravPackpaper}.

As an illustrative example, we will exhibit  how we obtain the  $\tilde \chi^0_i \, {\widetilde G} \, W^+ \, W^-$ interaction Lagrangian,
and the corresponding structure ``struc1" in Figure~\ref{fig:FAstructures1} used in our numerical calculation.
For this, we will need the term in the second line of Eq.~(\ref{eq:gravitino interaction lagrangian}) and 
from  Eq.~(\ref{eq:field strength tensor}) the part
that is proportional  to the $SU(2)_L $ gauge coupling constant $g$.
In particular,
we get $ {\l}^{(2)\, a} =  {\tilde W}^3$,   $f^{(2)\,abc} = \e^{abc}$ with $\e^{123} = 1$, and  ${A}_\m^{(2)\,1,2} = W_\m^{1,2}$. The
relevant Lagrangian can be written as
\begin{equation}
 \mathcal{L} =  \frac{i g}{8\mplanck}\overline{\widetilde G}_\m [\g^\n,\g^\r] \g^\m
{\tilde W}^3\,    \left( W^1_\n W^2_\r - W^2_\n W^1_\r \right)\, .  
\end{equation}
We write the field $ {\tilde W}^3$ in terms of the physical neutralino field as,
${\tilde W}^3  =  Z^*_{k,2} P_L \tilde\chi^0_k + Z_{k,2} P_R \tilde\chi^0_k$ and  the $W_\m^{1,2}$ 
components in terms of the physical $W$-boson fields as,
$W_\m^{1} =  (W^+_\mu + W^-_\mu)/\sqrt2$, $W_\m^{2} = i \, (W^+_\mu - W^-_\mu)/\sqrt2$.
Thus,  we get 
\begin{eqnarray}
{\cal L} & \sim & {g \over 4 \mplanck} W^+_\n W^-_\r  \overline{\widetilde G}_\m  [\g^\n,\g^\r] \g^\m
 (Z^*_{i,2} P_L + Z_{i,2} P_R) \tilde\chi^0_i \,. \label{vertexNGrWW}
\end{eqnarray}
Making use of the reality of the Lagrangian, from the hermitian conjugate part one gets
\begin{eqnarray}
{\cal L}  & \sim & {g \over 4 \mplanck} W^+_\n W^-_\r  \overline{\tilde\chi^0_i} \g^\m [\g^\n,\g^\r]
 (Z_{i,2} P_L + Z^*_{i,2} P_R)
{\widetilde G}_\m\, .
 \label{vertexNGrWWhc}
\end{eqnarray}
From this formula we calculate  the Feynman rule for the vertex in Figure~\ref{fig:vertexNeuGrWW}.
Finally, to obtain the {\tt FA} structure ``struc1",  given in Figure~\ref{fig:FAstructures1}, 
we write the vertices for an incoming and outgoing gravitino by adding the terms from
Eqs.~(\ref{vertexNGrWW}) and~(\ref{vertexNGrWWhc}) as
\begin{eqnarray}
&&
C[\tilde\chi^0_i, {\widetilde G}_\m, W^+_\nu, W^-_\rho]_{(1)}\, \g^\mu [\g^\nu,\, \g^\rho] P_L +
C[\tilde\chi^0_i, {\widetilde G}_\m, W^+_\nu, W^-_\rho]_{(2)}\, \g^\mu [\g^\nu,\, \g^\rho] P_R + \nonumber\\
&& C[\tilde\chi^0_i, {\widetilde G}_\m, W^+_\nu, W^-_\rho]_{(3)}\, [\g^\nu,\, \g^\rho] \g^\mu  P_L +
C[\tilde\chi^0_i, {\widetilde G}_\m, W^+_\nu, W^-_\rho]_{(4)}\, [\g^\nu,\, \g^\rho] \g^\mu  P_R\, ,
\end{eqnarray}
with the coupling vector
\begin{eqnarray}
C[\tilde\chi^0_i, {\widetilde G}_\m, W^+_\nu, W^-_\rho]_{(1)} & = &  i {g \over 4 \mplanck} Z_{i,2}\,,  \nonumber\\
C[\tilde\chi^0_i, {\widetilde G}_\m, W^+_\nu, W^-_\rho]_{(2)} & = &  i {g \over 4 \mplanck} Z^*_{i,2}\,,  \nonumber\\
C[\tilde\chi^0_i, {\widetilde G}_\m, W^+_\nu, W^-_\rho]_{(3)} & = &  i {g \over 4 \mplanck} Z^*_{i,2}\,  \nonumber\\
C[\tilde\chi^0_i, {\widetilde G}_\m, W^+_\nu, W^-_\rho]_{(4)} & = &  i {g \over 4 \mplanck} Z_{i,2}\, .
\end{eqnarray}
Note, that $C[\_, \_, V_{1\n}, V_{2 \r}]_{(i)} = -C[\_, \_, V_{2\n}, V_{1\r}]_{(i)}$.

Similarly, we have calculated all the other coefficient vectors $C[\ldots]$ for the ``struc1,2,3,4" shown 
in Figure~\ref{fig:FAstructures1}, in order to get the full list of all possible gravitino interactions in the MSSM.

\subsection{Two-body decays}

The gravitino decays into a fermion $F$, and a scalar $S$ or a vector particle $V$ are
\begin{eqnarray}
\widetilde G & \to & F  \,  S  \nonumber \\
& \to & f \tilde f^*_i,\, \tilde \chi^0_j   \,  (h^0, H^0, A^0),\, \tilde \chi^+_k  \,  H^-\,,  \nonumber \\
\widetilde G & \to & F  \,  V  \nonumber \\
& \to & \tilde g  \,  g,\, \tilde \chi^0_j  \,  (Z^0, \gamma),\, \tilde \chi^+_k   \,  W^-\,,
\label{two-body decays}
\end{eqnarray}
where $f$ denotes a SM fermion, quarks $(u,d,c,s,t,b)$ and  leptons $(e^-,\mu^-, \tau^-, \nu_e, \nu_\mu, \nu_\tau)$.
Its corresponding superpartners are the squarks $\tilde f_i$, $i = 1,2$. 
The four neutralino states are 
 $\tilde \chi^0_j$, $j = 1,\ldots,4$ and the two charginos 
$\tilde \chi^\pm_k$, $k = 1, 2$. With $g$ we denote the gluon and with $\tilde g$ its superpartner, the gluino.
Furthermore, in the MSSM we have
three neutral Higgs bosons (two $CP$-even:  $h^0$ and $H^0$, and one  $CP$-odd: $A^0$) and two charged 
Higgs bosons: $H^\pm$. The vector bosons are the photon $\gamma$, the $Z$-boson $Z^0$ and the $W$-bosons $W^\pm$.

The two-body matrix element ${\cal M}_2$ is auto-generated with the help of the {\tt FA}/{\tt FC} packages~\cite{FAref,FAFCref}. 
The decay width for the processes in Eq.~(\ref{two-body decays}) is given by
\begin{equation}
\Gamma(\widetilde G \to F X) = {N_c \over 16 \pi}\, {\kappa(m^2_{\widetilde G}, m^2_{F},  m^2_{X}) \over m^3_{\widetilde G}}\, {1 \over 4}\,
| {\cal M}_2|^2\, ,
\label{twobodydecaywidth}
\end{equation}
where $X = S,V$ and $N_c$ is the color factor ($3$ for $q\,\tilde q$,
$8$ for the $g\,\tilde{g}$ channel, and $1$ otherwise).
Please note the factor $ {1 \over 4}$ being the spin average of the incoming massive gravitino with spin~${3 \over 2}$. 
As usual, the kinematic function $\kappa$ is defined as  $\kappa^2( x, y, z) = (x - y - z)^2 - 4 y z$.
In parallel, we also calculated these two-body decay processes analytically and included them into {\tt GravitinoPack}.
This enables us to cross-check the auto-generated {\tt FA}/{\tt FC} code. 
Details will be published in \cite{GravPackpaper}.

\subsection{Three-body decays}

The gravitino decays into a neutralino and a pair of SM particles are
\begin{eqnarray}
\widetilde G & \to  & \tilde \chi_i^0 \, \bar f f\, ,  \nonumber \\
\widetilde G & \to  & \tilde \chi_i^0 \, V V\, ,\quad V V = Z^0 Z^0\,, Z^0 \gamma\,, W^+ W^- \, \nonumber \\
\widetilde G & \to  & \tilde \chi_i^0 \, V S\, ,\quad V S = (Z^0, \gamma)(h^0, H^0, A^0), W^+ H^-, W^- H^+\, \nonumber \\
\widetilde G & \to  & \tilde \chi_i^0 \, S S\, ,\quad S S = (h^0, H^0, A^0) (h^0, H^0, A^0), H^+ H^-\,.
\label{three-body decays}
\end{eqnarray}
These are 19 three-body  decay channels. All possible three-body decay channels for the process $\grav \to \tilde \chi^0_i \, X \, Y$ are given
in Table~\ref{tab:3bodydecays}.
It is worth to note that $\widetilde G \to \tilde \chi^0_i  \,  W^-  \,  H^+$ and its charge conjugated process
$\widetilde G \to \tilde \chi^0_i  \,  W^+  \,  H^-$ count as individual contributions, but
$\Gamma(\widetilde G \to \tilde \chi^0_i   \, W^-  \,  H^+) =  \Gamma(\widetilde G \to \tilde \chi^0_i  \,  W^+  \,  H^-)$.

\newcommand*{\myalign}[2]{\multicolumn{1}{#1}{#2}}
\renewcommand{\arraystretch}{1.3}
\begin{table}[t!]
\begin{center}
$\begin{array}{|l|c|l|l|}
\hline
\myalign{|c}{\rm process} & \myalign{|c}{\rm number} & \myalign{|c}{\rm first~decay} & \myalign{|c|}{\rm possible} \\[-2mm]
\myalign{|c}{\widetilde G \to \tilde \chi^0_i X Y} & \myalign{|c}{\rm of~graphs} & \myalign{|c}{\widetilde G \to \tilde X Y} & \myalign{|c|}{\rm resonances}\\
\hline 
\hline
\tilde \chi_i^0 f \bar f  & 7 & \tilde \chi_i^0 (h^0 , H^0 , A^0 , \gamma , Z^0 ) , f \tilde f^*_l , \tilde f_l \bar f  & h^0 , H^0 , A^0 , Z^0 , \tilde f_l \\ 
\tilde \chi_i^0 Z^0 Z^0  & 4 & \tilde \chi_i^0 (h^0 , H^0 ) , Z^0 \tilde \chi_k^0 , \tilde \chi_k^0 Z^0  & H^0 , \tilde \chi_k^0 \\ 
\tilde \chi_i^0 Z^0 \gamma  & 1 & \tilde \chi_k^0 \gamma  & \tilde \chi_k^0 \\ 
\tilde \chi_i^0 W^+ W^-  & 6 + 4pt  & \tilde \chi_i^0 (h^0 , H^0 , \gamma , Z^0 ) , W^+ \tilde \chi_j^- , \tilde \chi_j^+ W^-  & H^0 , \tilde \chi_j^\pm \\ 
\tilde \chi_i^0 Z^0 h^0  & 4 + 4pt  & \tilde \chi_i^0 (A^0 , Z^0 ) , Z^0 \tilde \chi_k^0 , \tilde \chi_k^0 h^0  & A^0 , \tilde \chi_k^0 \\ 
\tilde \chi_i^0 Z^0 H^0  & 4 + 4pt  & \tilde \chi_i^0 (A^0 , Z^0 ) , Z^0 \tilde \chi_k^0 , \tilde \chi_k^0 H^0  & A^0 , \tilde \chi_k^0 \\ 
\tilde \chi_i^0 Z^0 A^0  & 4 + 4pt  & \tilde \chi_i^0 (h^0 , H^0 ) , \tilde \chi_k^0 Z^0 , A^0 \tilde \chi_k^0  & H^0  , \tilde \chi_k^0 \\ 
\tilde \chi_i^0 \gamma h^0  & 1 & \tilde \chi_k^0 \gamma  & \tilde \chi_k^0 \\ 
\tilde \chi_i^0 \gamma H^0  & 1 & \tilde \chi_k^0 \gamma  & \tilde \chi_k^0 \\ 
\tilde \chi_i^0 \gamma A^0  & 1 & \tilde \chi_k^0 \gamma  & \tilde \chi_k^0 \\ 
\tilde \chi_i^0 W^+ H^-  & 5 + 4pt  & \tilde \chi_i^0 (h^0 , H^0 , A^0 ) , W^+ \tilde \chi_j^- , \tilde \chi_j^+ H^-  & H^0 , A^0 , \tilde \chi_j^\pm \\ 
\tilde \chi_i^0 W^- H^+  & 5 + 4pt  & \tilde \chi_i^0 (h^0 , H^0 , A^0 ) , W^- \tilde \chi_j^+ , \tilde \chi_j^- H^+  & H^0 , A^0 , \tilde \chi_j^\pm \\ 
\tilde \chi_i^0 h^0 h^0  & 4 & \tilde \chi_i^0 (h^0 , H^0 ) , h^0 \tilde \chi_k^0 , \tilde \chi_k^0 h^0  & H^0 , \tilde \chi_k^0 \\ 
\tilde \chi_i^0 H^0 H^0  & 4 & \tilde \chi_i^0 (h^0 , H^0 ) , H^0 \tilde \chi_k^0 , \tilde \chi_k^0 H^0  & \tilde \chi_k^0 \\ 
\tilde \chi_i^0 h^0 H^0  & 4 & \tilde \chi_i^0 (h^0 , H^0 ) , h^0 \tilde \chi_k^0 , \tilde \chi_k^0 H^0  &  \tilde \chi_k^0 \\ 
\tilde \chi_i^0 A^0 A^0  & 4 & \tilde \chi_i^0 (h^0 , H^0 ) , A^0 \tilde \chi_k^0 , \tilde \chi_k^0 A^0  & H^0 , \tilde \chi_k^0 \\ 
\tilde \chi_i^0 h^0 A^0  & 3 & \tilde \chi_i^0 (A^0 , Z^0 ) , h^0 \tilde \chi_k^0   & \tilde \chi_k^0 \\ 
\tilde \chi_i^0 H^0 A^0  & 3 & \tilde \chi_i^0 (A^0 , Z^0 ) , H^0 \tilde \chi_k^0  & \tilde \chi_k^0 \\ 
\tilde \chi_i^0 H^+ H^-  & 6 & \tilde \chi_i^0 (h^0 , H^0 , \gamma , Z^0 ) , H^+ \tilde \chi_j^- , \tilde \chi_j^+ H^-  & H^0 , \tilde \chi_j^\pm \\ 
\hline  
\end{array}$
\end{center}
\renewcommand{\arraystretch}{1}
\caption[3body decay table]{All possible three-body decays 
channels of the gravitino into a neutralino and a pair of SM particles; $4pt$ denotes a Feynman graph
with four-point interaction, like graph a) in Figure~\ref{fig:eynGr2NeuWmWp}. The indices are $i = 1, \ldots, 4$; $k = 2,3,4$; $j,l = 1,2$.}
\label{tab:3bodydecays}
\end{table}

As before, the three-body matrix element ${\cal M}_3$ is auto-generated with the help of the {\tt FA}/{\tt FC} packages. 
The corresponding four-momenta for the decay 
$\grav \to \tilde \chi^0_i \, X \, Y$ are fixed by the relation $k_1 = k_2 + k_3 + k_4$, where $X$ and $Y$ are given in 
Eq.~(\ref{three-body decays}). 
In our code, the calculation can be performed in three different center of mass systems. This option
was used as an additional check for our numerical results.
For illustration, we show here only the particular system fixed by the kinematical relation: $\vec k_1 - \vec k_2 = \vec q = \vec k_3 + \vec k_4 = 0$.
The Mandelstam variables $M_1^2$ and $T$ are defined as 
\begin{eqnarray}
(k_1 - k_2)^2  & =  M_1^2 =  & (k_3 + k_4)^2\, ,\nonumber\\
(k_1 - k_3)^2  & = T =  & (k_2 + k_4)^2\, .
\end{eqnarray}
The decay width  
for the processes in Eq.~(\ref{three-body decays}) is given in differential form by
\begin{equation}
{{\rm d}^2 \Gamma(\widetilde G \to\tilde \chi_i^0  \, X  \,  Y) \over {\rm d} M_1^2\, {\rm d} T} = 
{N_c \over 256  \,  \pi^3  \,  m^3_{\widetilde G}}\, {1 \over 4}\,  |{\cal M}_3|^2\,.
\end{equation} 
With $p^* = \kappa(M_1^2, m^2_{\widetilde G}, m^2_{\tilde \chi_i^0})/(2 M_1)$ and
$k^* = \kappa(M_1^2, m^2_X, m^2_Y)/(2 M_1)$,
$E^2_1 = m^2_{\widetilde G} + p^{* 2}$  and $E^2_3 = m^2_{X} + k^{* 2}$  we can write the lower and upper 
bound of  $T$ as 
\begin{equation}
T^{\rm min, max} = m^2_{\widetilde G}  +  m^2_X  - 2 E_1 E_3 \mp 2 p^* k^*  \, .
\end{equation}
We get 
\begin{equation}
 \Gamma(\widetilde G \to\tilde \chi_i^0 X Y) = 
{N_c \over 256  \,  \pi^3  \,  m^3_{\widetilde G}}\, {1 \over 4}\,  \int_{(m_X + m_Y)^2}^{(m_{\widetilde G} - m_{\tilde \chi^0_i})^2} {\rm d}M_1^2 
 \int_{T^{\rm min}}^{T^{\rm max}} {\rm d}T \, |{\cal M}_3|^2\,.
 \label{threebodydecaywidth}
\end{equation} 

\begin{figure}[t!]
\begin{center}
\resizebox{14cm}{!}{\includegraphics{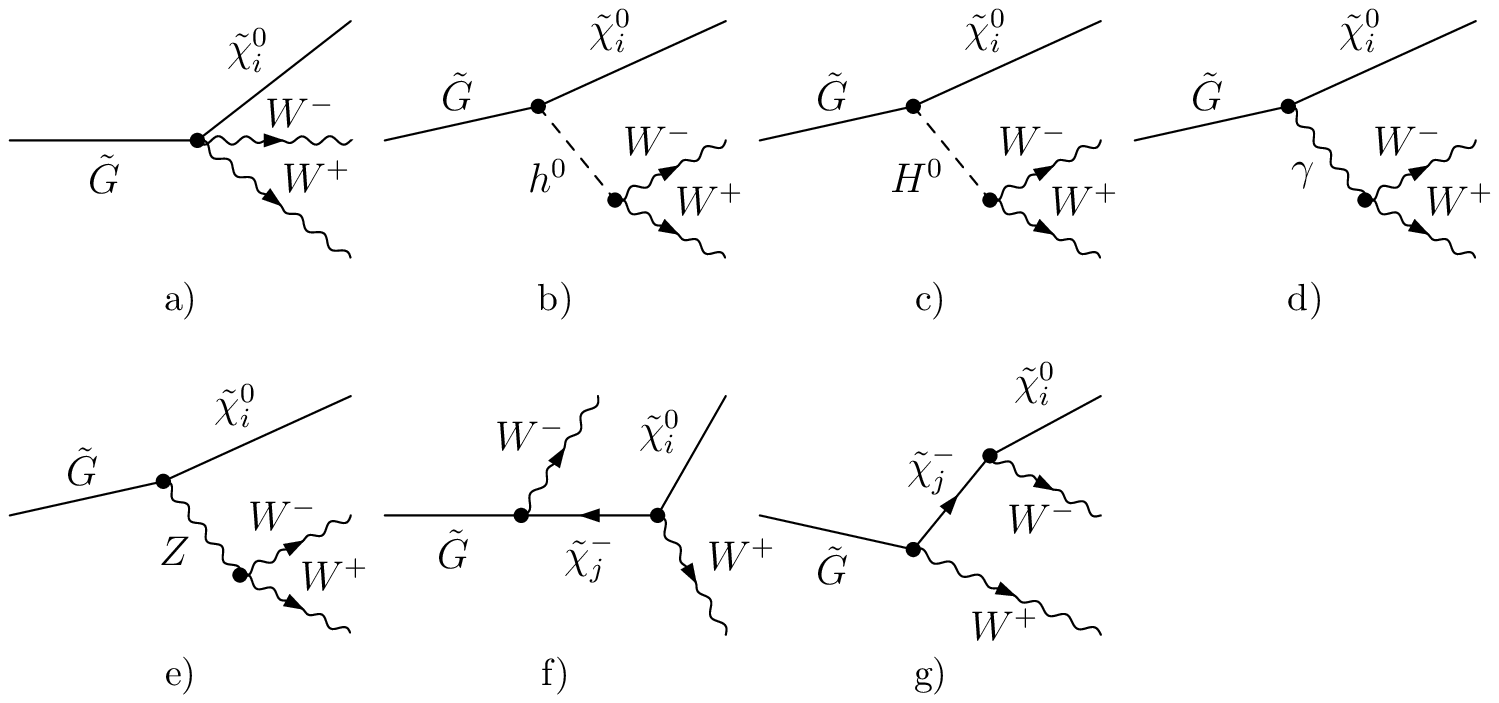}}
\end{center}
\caption{Feynman graphs for the decay ${\widetilde G} \to \tilde \chi^0_i W^- W^+$.
\label{fig:eynGr2NeuWmWp}}
\end{figure}

In general, the matrix element ${\cal M}_3$ is a sum of different amplitudes. 
As in the subsection~\ref{subsec:inter}, we are using the process  ${\widetilde G} \to \tilde \chi^0_i  \,  W^-  \,  W^+$ 
as an example to demonstrate how one calculates the resonant, the non-resonant and the interference part of  
Eq.~ \ref{threebodydecaywidth}.
For this particular process there exist 7~Feynman graphs, plotted in  Figure~\ref{fig:eynGr2NeuWmWp},  which are 9 individual amplitudes.
In this case only the $H^0$ and $\tilde \chi^\pm_{1,2}$ propagators can become resonant.
We assume that we have  $H^0$  resonant in ${\cal M}^{c)}_3$:
$m_{\widetilde G} > m_{\tilde \chi_i^0} + m_{H^0}$ and $m_{H^0} > 2 m_W$.
We write ${\cal M}_3$ as
\begin{equation}
{\cal M}_3 = \sum_{i \ne c}  {\cal M}^{i)}_3 +  {\cal M}^{c)}_3 = {\cal M}^{{\rm rest}}_3 +  
{\cal M}_{2,1} (M_1^2) \, \,   {i \over M_1^2 - m^2_{H^0} + i m_{H^0} \Gamma_{H^0} } \, \,    {\cal M}_{2,2} (M_1^2, T)
\end{equation}
with ${\cal M}_{2,1} =  {\cal M}(\widetilde G \to  \tilde \chi_i^0 + H^0)$ and ${\cal M}_{2,2} =  {\cal M}(H^0 \to W^- W^+)$
and $\Gamma_{H^0}$ is the total decay width of $H^0$. 
We get
\begin{eqnarray}
|{\cal M}_3|^2 
      & = &  |{\cal M}^{{\rm rest}}_3|^2 +  |{\cal M}_{2,1}|^2 \,  | \, {\cal M}_{2,2}|^2  \, \, 
                     {1 \over (M_1^2 - m^2_{H^0})^2 +  m^2_{H^0} \Gamma^2_{H^0}} \nonumber\\
&& + 2 {\rm Re}\left(({\cal M}^{{\rm rest}}_3)^*  \, {\cal M}_{2,1}  {\cal M}_{2,2} \, 
                                            \,      {i (M_1^2 - m^2_{H^0}) + m_{H^0} \Gamma_{H^0} \over
(M_1^2 - m^2_{H^0})^2 +  m^2_{H^0} \Gamma^2_{H^0} } \right) \, .
 \label{threebodydecayexample}
\end{eqnarray}
We use
\begin{equation}
{\rm lim}_{m \Gamma \to 0}\left( {1 \over (s - m^2)^2 + m^2 \Gamma^2}\right) = {\pi \over m \Gamma} \,  \delta(s - m^2)
\end{equation}
and apply the narrow width approximation (NWA) for $H^0$,  $\Gamma_{H^0} \ll m_{H^0}$. We have three terms
in Eq.~(\ref{threebodydecayexample}). The first one, $\Gamma^{\rm non-reso}$, leads to the 
remaining non-resonant three-body decay width. 
To get this, we just have to substitute ${\cal M}_3$ in Eq.~(\ref{threebodydecaywidth}) by 
${\cal M}^{{\rm rest}}_3$. The second one corresponds to  the resonant part. Using ${\rm d}T = 2 p^* k^* {\rm d}\cos\theta$ 
and $M_1^2 \to m^2_{H^0}$ we get
\begin{equation}
 \Gamma^{\rm reso}(\widetilde G \to\tilde \chi_i^0  \,  W^-  \,  W^+) = 
 {1 \over 4}\, {1 \over 128 \, \pi^3 \, m^3_{\widetilde G}}\,  {\pi \over  m_{H^0}  \Gamma_{H^0} }\, \,  p^* k^* |{\cal M}_{2,1}|^2
 \int_{-1}^{1} {\rm d}\cos\theta  |{\cal M}_{2,2}|^2  \, .
 \label{threebodydecaywidthres}
\end{equation} 
${\cal M}_{2,2}$ must be independent on $\cos\theta$ because $H^0$ is at rest. Using Eq.~(\ref{twobodydecaywidth}) we have
 \begin{equation}
\Gamma(\widetilde G \to \tilde \chi^0_i H^0) =  {1 \over 4} {m_{H^0} \over 8 \pi}\, {p^* \over m^3_{\widetilde G}}\,  |{\cal M}_{2,1}|^2
\quad {\rm and} \quad  
\Gamma(H^0 \to W^- W^+) = {1 \over 8 \pi}\, {k^* \over m^2_{H^0}}\,  |{\cal M}_{2,2}|^2\, .
 \end{equation}
Note that $(m_{H^0}/m_{\widetilde G}) \, p^* = \kappa(m^2_{\widetilde G},m^2_{H^0}, m^2_{\chi^0_i})/(2 m_{\widetilde G})$ is the 
momentum of the outgoing particles in the restframe of the gravitino.
We can write Eq.~(\ref{threebodydecaywidthres}) as
\begin{equation}
 \Gamma^{\rm reso}(\widetilde G \to\tilde \chi_i^0 W^- W^+) =  \Gamma(\widetilde G \to \tilde \chi^0_i H^0) \, {\Gamma(H^0 \to W^- W^+) \over \Gamma_{H^0}}\, .
 \label{threebodydecaywidthres1}
\end{equation} 
The third term in Eq.~(\ref{threebodydecayexample}) denotes the interference. It can be written in the NWA as
\begin{equation}
 \Gamma^{\rm if}(\widetilde G \to\tilde \chi_i^0  \,  W^-  \,  W^+) = 
 {1 \over 4}\,\, {1 \over 64 \pi^2 m^3_{\widetilde G}}\,\, p^* k^*  \int_{-1}^{1} {\rm d}\cos\theta \, 
 {\rm Re}\left(({\cal M}^{{\rm rest}}_3)^* {\cal M}_{2,1}  {\cal M}_{2,2} \right)  \, .
 \label{threebodydecaywidthinterference}
\end{equation} 

We can generalize our formulas to include $n$ resonances, for example, see the last column in Table~\ref{tab:3bodydecays}.
The total matrix element is a sum of non-resonant and a sum of resonant matrix elements. The sum of the non-resonant matrix elements is denoted
as before as ${\cal M}^{{\rm rest}}_3$. 
Moreover, we split the sum of the resonant matrix elements, ${\cal M}^{n}_3$,   into the matrix element with
the particular resonance $j$, ${\cal M}^{j}_3$, and the other ones, ${\cal M}^{!j}_3 =  \sum_{k \ne j} {\cal M}^{k}$.
The result for the non-resonant part is the same like in the case with only one resonance. The resonant part is
\begin{equation}
 \Gamma^{\rm reso}(\widetilde G \to\tilde \chi_i^0 X Y) =
 {1 \over 512\, \pi^2\, m^3_{\widetilde G}}\,   \sum_{j}\, {p^* k^*  \over  m_{j} \Gamma_{j} }\, |{\cal M}^j_{2,1}|^2
 \int_{-1}^{1} {\rm d}\cos\theta  |{\cal M}^j_{2,2}|^2  \, ,
 \label{threebodydecaywidthresN}
\end{equation}
and the interference term reads
\begin{equation}
 \Gamma^{\rm if}(\widetilde G \to\tilde \chi_i^0 X Y) = 
 {1 \over 256\, \pi^2\, m^3_{\widetilde G}}\,  \sum_{j}\, p^* k^*  \int_{-1}^{1} {\rm d}\cos\theta \, 
 {\rm Re}\left[\left( {\cal M}^{{\rm rest}}_3 +   {\textstyle {1 \over 2}}  {\cal M}^{!j}_3\right)^*  {\cal M}^j_{2,1}  {\cal M}^j_{2,2} \right]  \, .
 \label{threebodydecaywidthinterferenceN}
\end{equation} 
The factor 1/2 in front of ${\cal M}^{!j}_3$ is necessary in order to avoid double counting of the interference terms
within $|{\cal M}^{n}_3|^2$. These NWA formulas, Eqs.~(\ref{threebodydecaywidthresN})~and~(\ref{threebodydecaywidthinterferenceN}),
are only applicable when there is no overlap of resonances, 
$\Gamma_j + \Gamma_k \ll |m_j - m_k|$ for all possible couples of $j$ and $k$.

\section{Numerical results}
\label{sec:numerics}

In our numerical analysis we choose representative points from various supersymmetric models, assuming 
different  mechanism for  the supersymmetry breaking. In  particular, we study points based 
on the pMSSM~\cite{pMSSM},   the 
CMSSM~\cite{cmssm1,cmssm2} model and the  NUHM~\cite{nuhm}. 
These points satisfy the sparticle mass bounds from LHC~\cite{LHC} and the cosmological constraints on 
the amount of the dark matter as measured by the WMAP ~\cite{WMAP9} and 
Planck~\cite{planck}  experiments, assuming that the dark matter is made of neutralinos.
In addition, we use the branching ratio for the process $B_s \to \mu^+ \mu^-$
as it is measured from LHCb and other experiments~\cite{bmm,LHCb}, Higgs mass at $\sim 126$~GeV \cite{mh125}, and the recent XENON100~ \cite{xenon100} direct detection bound.
\begin{table}[h!]
\begin{center}
\begin{tabular}{c|c|c}
\hline 
  Parameters  & Coannihilation point & $A^0$-funnel point \\ 
\hline  \hline 
       $\tan\beta=  {\langle H^0_2  \rangle}/ {\langle H^0_1  \rangle}$ & 30  &  20 \\ 
       $\mu$,  Higgsino mixing parameter &  2200 GeV  & 700 GeV     \\ 
       $M_A$,  $A^0$ Higgs boson mass & 1100 GeV &   770 GeV     \\ 
       ($M_1$, $M_2$, $M_3$),  Gauginos  masses& (399,1000,3000) GeV & (400,800,2400) GeV \\ 
    $A_t$,  top  trilinear coupling & $-2300$ GeV &   $-2050$ GeV      \\ 
    $A_b$,  bottom  trilinear coupling &  $-2300$ GeV &   $-2050$ GeV      \\ 
    $A_\tau$, tau  trilinear coupling & $ -2300$ GeV &  $-1000$ GeV      \\ 
       $m_{\widetilde{q}_{L}}$,  1st/2nd family  $Q_L$ squark mass & 1500 GeV &  1500 GeV   \\ 
       $m_{\widetilde{u}_{R}}$,  1st/2nd family  $U_R$ squark &  1500 GeV &   1500 GeV    \\  
       $m_{\widetilde{d}_{R}}$,   1st/2nd family $D_R$ squark &  1500 GeV &  1500  GeV    \\  
       $m_{\widetilde{\ell}_{L}}$,   1st/2nd family  $L_L$ slepton & 800 GeV &  600   GeV    \\ 
       $m_{\widetilde{e}_{R}}$,   1st/2nd family  $E_R$ slepton & 2500 GeV &    2500  GeV  \\  
      $m_{\widetilde{Q}_{3L}}$,  3rd family  $Q_L$ squark & 1100 GeV  &       800    GeV      \\  
      $m_{\widetilde{t}_{R}}$,  3rd family  $U_R$ squark & 1100 GeV &       1000    GeV          \\  
       $m_{\widetilde{b}_{R}}$,  3rd  family $D_R$ squark & 1100 GeV &   1500     GeV          \\
       $m_{\widetilde{L}_{3L}}$,  3rd  family $L_L$ slepton & 400 GeV &  800     GeV               \\ 
       $m_{\widetilde{\tau}_{R}}$,  3rd  family $E_R$ slepton & 2000 GeV &      2000  GeV       \\ 
\hline         
\end{tabular}
\end{center}
\caption{The pMSSM parameters used as  input for the two points studied  in our analysis.}
\label{tab:pMSSM}
\end{table}

As discussed in the previous section, the total three-body decay result consists of three parts:
the resonant part calculated, the non-resonant part, and the interference term.
The resonant part, naturally, vanishes below the threshold of the on-shell production of the intermediate particle.
The interference term is numerically always very small in the studied scenarios and negligible compared 
to the non-resonant and resonant parts. Therefore, we will not show explicitly this term in the figures with three-body gravitino
decays.
 
First we study two pMSSM points, one in the stau-coannihilation
region ($m_{\tilde{\tau}_1} \simeq m_{\tilde{\chi}_1^0}$), and the other one in the $A^0$-funnel region ($m_{A^0} \simeq 2 m_{\tilde{\chi}_1^0}$),
with the input parameters given in Table~\ref{tab:pMSSM}.
At the former point, we get a neutralino dark matter relic density compatible with WMAP data by increasing the neutralino pair
annihilation cross-section by adding $\tilde\chi_1^0$-$\tilde{\tau}_1$ coannihilation processes, while at the latter one
by using the rapid pair annihilation of the $\tilde{ \chi}_1^0$'s through the 
$A^0$ resonant exchange. 
For the coannihilation point we get the masses 
$m_{{\tilde{\chi}}^0_{1,2,3,4}}=(399,998,2201,2203) $~GeV, $ m_{\tilde{\chi}_{1,2}^+} = (998 ,\, 2203)$~GeV,
and $m_{H^+}=1103$ GeV. 
For the funnel point we get
$m_{{\tilde{\chi}}^0_{1,2,3,4}}=(397,675,703,830) $~GeV, $ m_{\tilde{\chi}_{1,2}^+} = (674 ,\, 830)$~GeV,
and $m_{H^+}=774$ GeV. 
These values will help us understand the various features appearing in the figures be discussed below.

\begin{figure}[t!]
\vspace*{0.3cm}
\begin{center}
\mbox{\includegraphics[width=8cm]{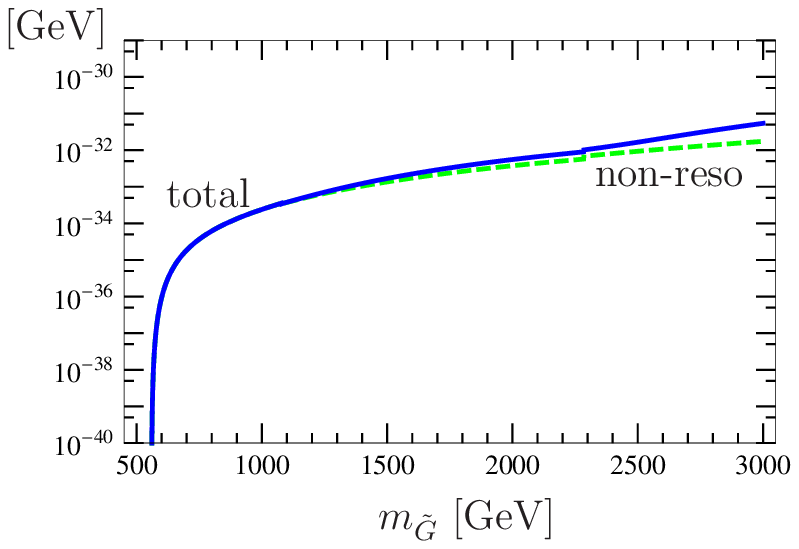} \hfill
\includegraphics[width=8cm]{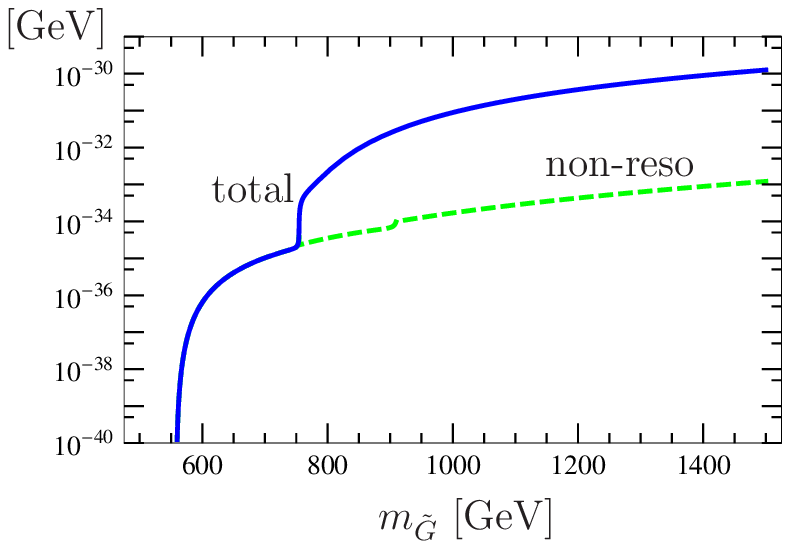}}
\end{center}
\vspace*{-0.5cm}
\caption[]{The decay width for the process $\grav \to  \lsp \, W^+ \, W^- $  left (right),
for the stau-coannihilation  ($A^0$-funnel) pMSSM point as discussed in the text. 
The solid blue line is the total width (resonant and non-resonant part) of this decay, while the dashed green line is the non-resonant part only.}
\label{fig:ww}  
\end{figure}

In Figure~\ref{fig:ww} left, the decay width for the processes $\grav \to  \lsp \, W^+ \, W^- $ is shown.
The solid blue line is the total width (resonant and non-resonant part), while the dashed green line is the non-resonant part only. 
As can be seen in Table~\ref{tab:3bodydecays}, for this three-body decay channel, we can have a resonant behavior through
the chargino ($\tilde{\chi}_{1,2}^+$) or $H^0$ exchange. This means, that for a gravitino mass
$ \mgrav= m_{\tilde{\chi}^{+}_j}  + M_W$ and $ \mgrav=m_{{\tilde{\chi}^0_{1}}} + m_{H^0} $, one expects to see a structure, 
being proportional to the resonant contribution as given in Eq.~(\ref{threebodydecaywidthresN}), following the NWA. 
The size of the NWA term, for example in the case of the ${\tilde{\chi}^{+}_1}$ resonance, 
 is proportional to the partial width $\tilde{\chi}^{+}_1  \to \tilde{\chi}^{0}_1 \, W^+$, and 
inversely proportional to the total width of ${\tilde{\chi}^{+}_1}$. In Figure~\ref{fig:ww} left,
there are very small structures that 
correspond to chargino thresholds $ \mgrav= m_{\tilde{\chi}^{+}_{1,2}}  + M_W$.
For this case the total chargino widths are $ \Gamma_{\tilde{\chi}^{+}_{1,2}} =(8,144)$~GeV.
The partial chargino width to $\tilde{\chi}^{0}_1 \, W^+$ is of the order $10^{-4}$ GeV
because $\chi^0_1$ is almost a pure bino state, and for a chargino-neutralino-$W$ interaction 
one needs a wino-wino or a higgsino-higgsino combination of chargino-neutralino.
Thus the effects are suppressed in this scenario. 
In Figure~\ref{fig:ww} right, we show the corresponding widths for the $A^0$-funnel point. 
For this point we have the total widths $ \Gamma_{\tilde{\chi}^{+}_{1,2}} =(0.7 ,\, 5)$ GeV, and
the partial width $\tilde{\chi}^{+}_1  \to \tilde{\chi}^{0}_1 \, W^+$ is almost three orders of magnitude bigger ($\mathcal{O}(0.1)$ GeV) because 
the charginos are mixed states of gauginos and higgsinos and $\chi_1^0$ is again a bino state but with non-negligible admixtures
of higgsinos and winos. This makes the partial gravitino width quite large after the $\tilde{\chi}^{+}_1 W^-$~threshold.

Before we discuss other gravitino decay channels some comments are in order. First, it is important to note 
that below the first threshold (the $\tilde{\chi}^{+}_1$ for this case) the knowledge of the full three-body amplitude is 
important in the computation of the gravitino decay width. Looking at Figure~\ref{fig:ww}  one notices that the size of the 
width  below and above the threshold is comparable, especially for the coannihilation point. Thus, using only the two body
decays (or the NWA terms) one gets, in general,  inadequate  results.   A second comment is related to the 
relative size of the resonant and the non-resonant contributions. If the partial width of the 
intermediate unstable particle, like the $\tilde{\chi}^{+}_1 \to \tilde{\chi}^{0}_1 \, W^+$ at the coannihilation point, is small these two contributions
are comparable. In this case the non-resonant contribution dominates and therefore
in Figure~\ref{fig:ww} left the blue and the green curves are very close. 
For example, in Figure~\ref{fig:ww}~left at $m_{\widetilde G} = 2$~TeV we get 
$\Gamma^{\rm reso} = 1.8 \times 10^{-33}$~GeV and $\Gamma^{\rm non-reso} = 3.8 \times 10^{-33}$~GeV.
This underlines the relevance of the full three-body calculation.

\begin{figure}[t!]
\begin{center}
\mbox{\includegraphics[width=8cm]{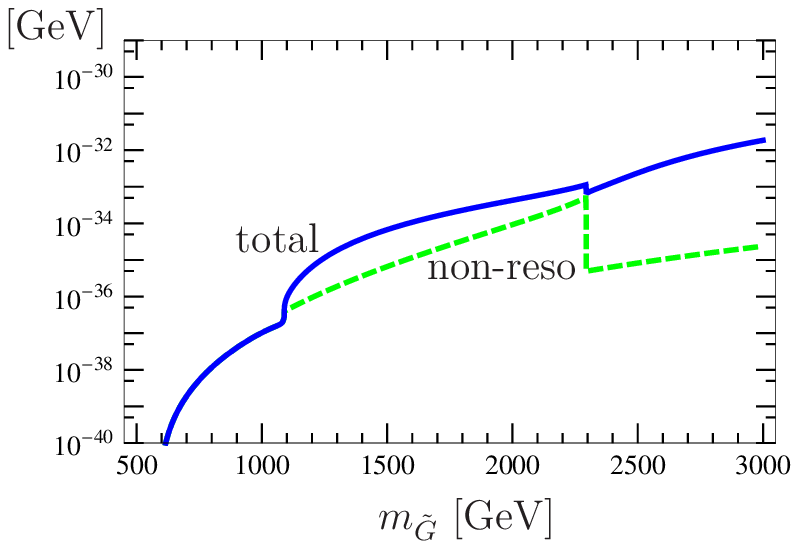} \hfill
\includegraphics[width=8cm]{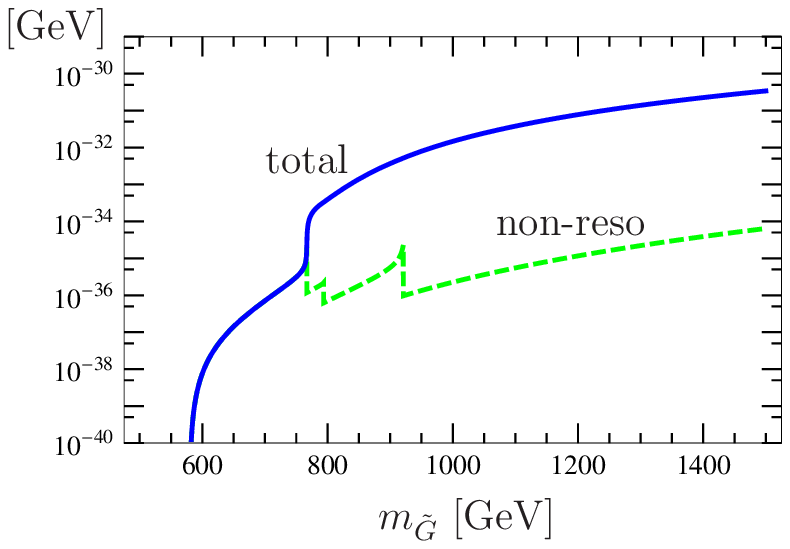}}
\end{center}
\vspace*{-0.5cm}
\caption[]{The decay width for the process $\grav \to  \lsp \, Z^0 \, Z^0 $  left (right),
for the stau-coannihilation  ($A^0$-funnel) pMSSM point as discussed in the text. The lines are as in Figure~\ref{fig:ww}.}
\label{fig:zz}  
\end{figure}

We now turn to Figure~\ref{fig:zz}, where we display the width for the  channel  $\grav \to  \lsp \, Z^0 \, Z^0 $.
In this case, the possible resonant particles, as can be seen in Table~\ref{tab:3bodydecays}, are the 
heavier neutralinos  $\tilde{\chi}_{2,3,4}^0$ and the heavy $CP$-even Higgs boson $H^0$.
The first kink in Figure~\ref{fig:zz} left (coannihilation point) corresponds to the threshold of the two body process $\grav \to \tilde{\chi}_2^0 \, Z^0 $,
where $ m_{{\tilde{\chi}}^0_{2}}=998  $ GeV. 
The corresponding increase in  Figure~\ref{fig:zz} right, for $ m_{{\tilde{\chi}}^0_{2}}=675  $ GeV, is much stronger because the 
partial width for the channel $\tilde{\chi}^{0}_2  \to \tilde{\chi}^{0}_1 \, Z^0$ is three order of magnitudes bigger, while 
the total width $ \Gamma_{\tilde{\chi}^{0}_{2}} $ is almost ten times smaller.  
For a neutralino-neutralino-$Z$ interaction both neutralino states must have a higgsino component. As already stated, in 
the coannihilation scenario the lightest neutralino is almost a pure bino state which 
suppresses the $\tilde{\chi}^{0}_2  \to \tilde{\chi}^{0}_1 \, Z^0$ coupling.
In addition, the effects of the thresholds of the neutralinos ${{\tilde{\chi}}^0_{3,4}}  $ with masses 702 and 829 GeV (see Figure~\ref{fig:zz} right), 
can be seen also in the dashed line denoting the non-resonant contribution.
 
\begin{figure}[t!]
\begin{center}
\mbox{\includegraphics[width=8cm]{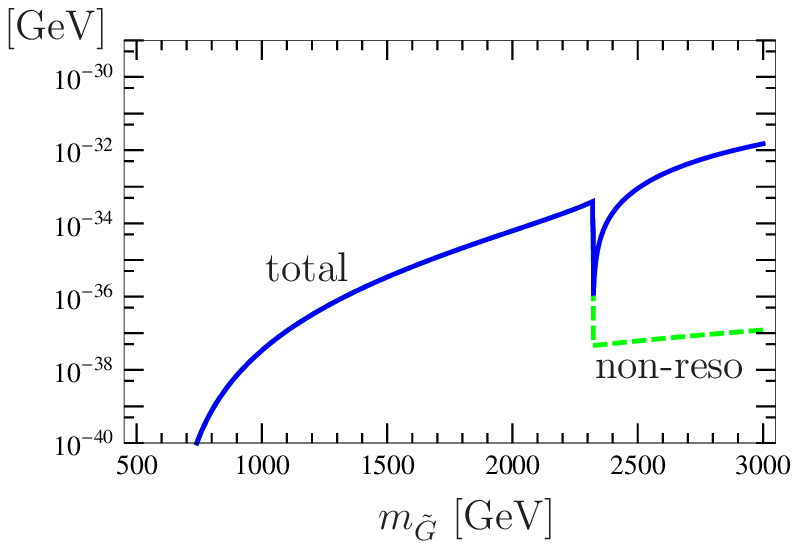} \hfill
\includegraphics[width=8cm]{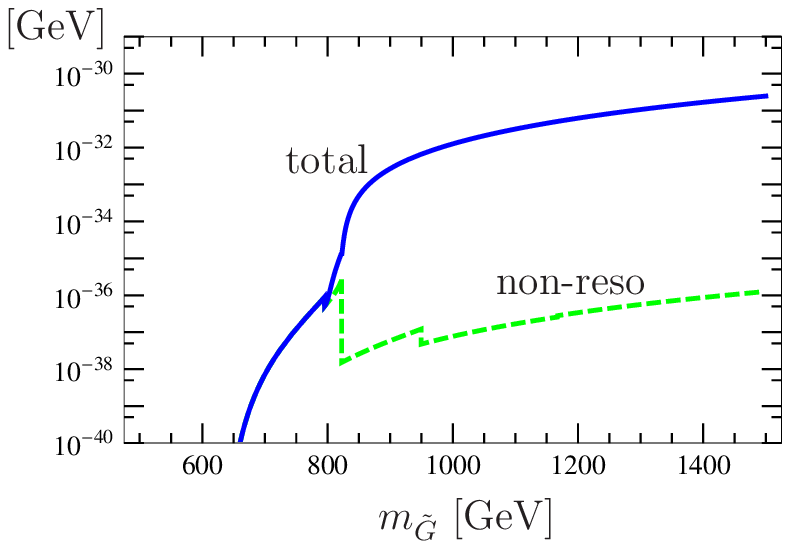}}
\end{center}
\vspace*{-0.5cm}
\caption[]{The decay width for the process $\grav \to  \lsp \, h^0 \, h^0 $  left (right),
for the stau-coannihilation  ($A^0$-funnel) pMSSM point as discussed in the text. The lines are as in Figure~\ref{fig:ww}.}
\label{fig:h0h0}  
\end{figure}

In Figure~\ref{fig:h0h0}, we study the channel  $\grav \to  \lsp \, h^0 \, h^0 $. 
The possible resonant exchanged particles are the same as in $\grav \to  \lsp \, Z^0 \, Z^0 $, namely the 
heavier neutralinos  $\tilde{\chi}_{2,3,4}^0$ and the heavy $CP$-even Higgs boson $H^0$.
In the coannihilation point (left) the effect of the enhancement after the threshold of the $\tilde{\chi}^{0}_2$ resonance 
at $m_{\widetilde G} \sim  1100$~GeV is
quite suppressed, because the $\tilde{\chi}^{0}_2  \to \tilde{\chi}^{0}_1 \, h^0$ partial width is small as the  $\tilde{\chi}^{0}_2  \to \tilde{\chi}^{0}_1 \, Z^0$
before.
 On the other hand, the $\tilde{\chi}^{0}_{3,4}  \to \tilde{\chi}^{0}_1 \, h^0$ widths are much bigger, producing this large structure in the gravitino 
decay width across the almost degenerate $\tilde{\chi}^{0}_{3,4}$ thresholds at $m_{\widetilde G} \sim  2300$~GeV.  
At the $A^0$-funnel point the situation is different. 
The threshold effect across the $\tilde{\chi}^{0}_{2}$ resonance at $m_{\widetilde G} = 801$~GeV  is much bigger due to the larger 
$\tilde{\chi}^{0}_2  \to \tilde{\chi}^{0}_1 \, h^0$ decay width. Moreover, the thresholds of the heavier 
neutralinos are also visible.

\begin{figure}[t!]
\begin{center}
\mbox{\includegraphics[width=8cm]{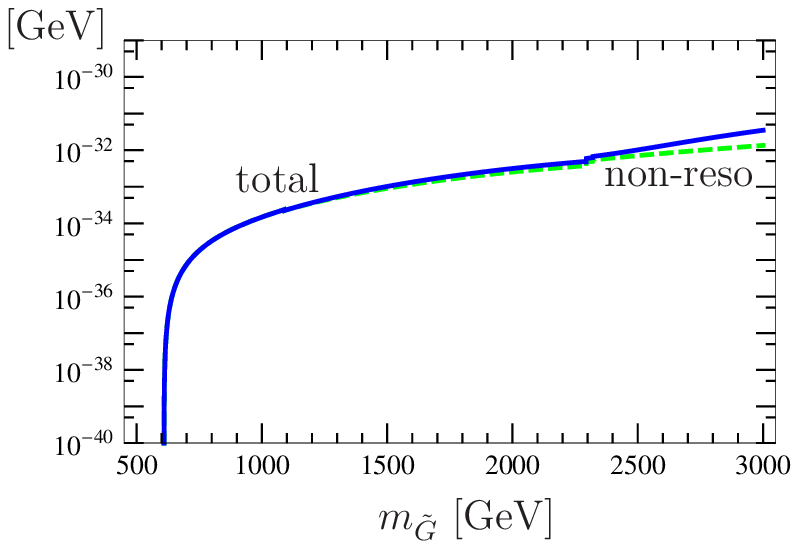} \hfill
\includegraphics[width=8cm]{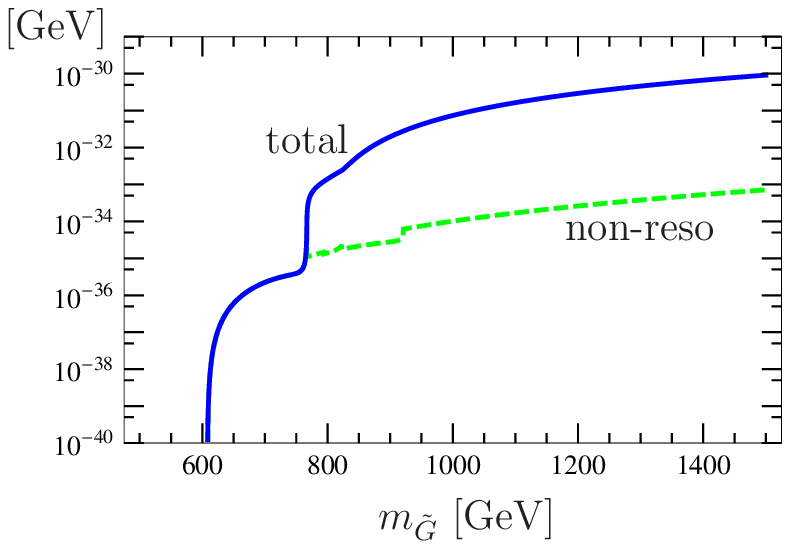}}
\end{center}
\vspace*{-0.5cm}
\caption[]{The decay width for the process $\grav \to  \lsp \, Z^0 \, h^0 $  left (right),
for the stau-coannihilation  ($A^0$-funnel) pMSSM point as discussed in the text. The lines are as in Figure~\ref{fig:ww}.}
\label{fig:Zh0}  
\end{figure}

Figure~\ref{fig:Zh0} shows the 
width for the decay $\grav \to  \lsp \, Z^0 \, h^0 $ as a function of $m_{\widetilde G}$. As in Figure~\ref{fig:ww}, the effect of the 
thresholds is not important for the coannihilation point (left). In contrast, especially the $\tilde{\chi}^{0}_{2} $ threshold  
can be seen at the $A^0$-funnel point (right). 
Note again, that at the coannihilation point (left) $\tilde{\chi}^{0}_{1}$ is a pure bino state. Therefore, the decays 
$ \tilde{\chi}^{0}_{2,3,4} \to \tilde{\chi}^{0}_{1} \, Z^0$ are suppressed and the non-resonant contribution is dominant.

\begin{figure}[t!]
\mbox{\hspace{2mm}
\includegraphics[width=7.7cm]{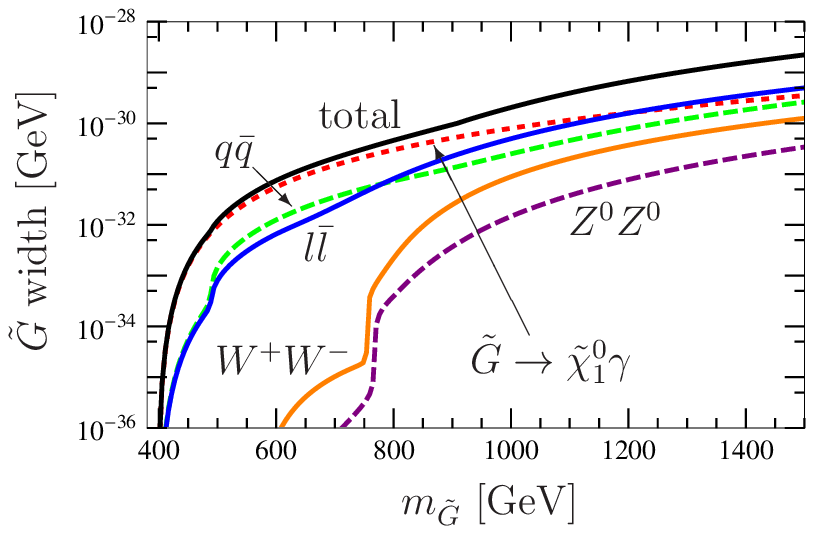}
\hspace{3mm}
\hfill
\includegraphics[width=7.6cm]{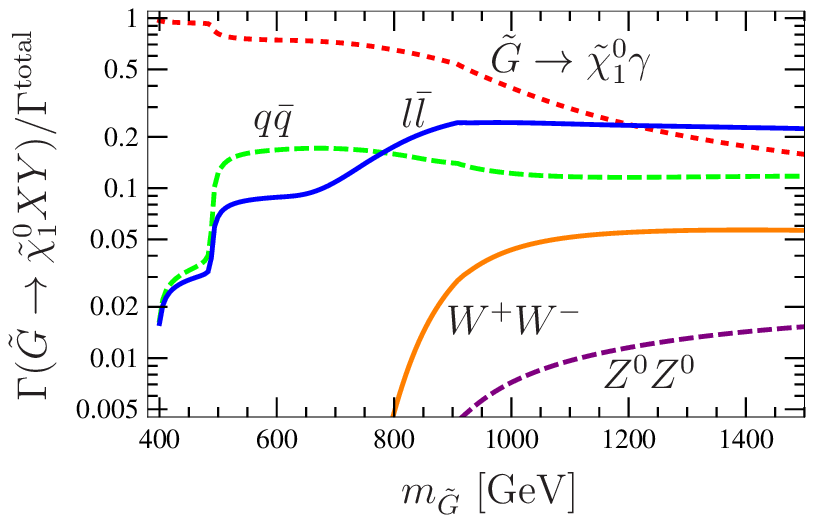}
}
\vspace*{-0.5cm}
\caption[]{The three-body decay widths of the gravitino decaying into $\lsp$ and $q \bar q$  ,$l \bar l$, $W$-pairs, and $Z$-pairs at the $A^0$-funnel point;
``total" denotes $\Gamma^{\rm total}$ which is the full two-body width plus the sum of the non-resonant part of three-body decay widths with $\lsp$;
$q \bar q$ stands for the sum over all six quark flavors and $l  \bar l$ for the sum over the three charged lepton and three neutrino flavors.
The red dotted lines denote the two-body decay $\widetilde G \to \lsp \gamma$.  In the right figure we display the corresponding branching ratios for 
the decay channels plotted in the left figure.}
\label{fig:NeuDec_funnel}
\end{figure}

Closing the discussion on the gravitino widths at these two representative pMSSM points, we show in Figure~\ref{fig:NeuDec_funnel}
the three-body decay widths of the gravitino decaying into $\lsp$ together with $q \bar q$, $l \bar l$, and $W, \,Z$-boson pairs at the $A^0$-funnel point.
We also show the two-body decay channel $\widetilde G \to \lsp \gamma$ as a reference, because it is dominant for small $m_{\widetilde G}$.
In the left figure we also show $\Gamma^{\rm total}$ which is the full two-body width plus the sum of the non-resonant part of 
three-body decay widths with $\lsp$,
denoted by ``total". In the right figure we show the relative quantities in terms of $\Gamma^{\rm total}$;
$q \bar q$ stands for the sum over all six quark flavors, $\sum_{i = 1,6} \Gamma^{\rm reso~+~non-reso }(\widetilde G \to \lsp q_i \bar q_i)$ and $l \bar l$,
$q_i = u, d, c, s, t, b$, and $l  \bar l$ the sum of the three charged lepton and three neutrino flavors, $\sum_{j = 1,3} \left(\Gamma^{\rm non-reso~+~reso}(\widetilde G \to \lsp l^+_j  l^-_j)\right.$ + $\left.\Gamma^{\rm non-reso~+~reso}(\widetilde G \to \lsp \nu_{l_j} \bar \nu_{l_j})\right) $, $l_j = e, \nu, \tau$. 
The decay width summing up the decays into all fermion pair can reach 38\%, into the $W$-boson pair 5.6\% and 
into the $Z$-boson pair 1.5\% in the shown range.
The analogous plots for the coannhilation point look similar but the decays into $W$- and $Z$-boson pairs are suppressed by about two orders of magnitude
due to the pure bino state of the LSP as already discussed before.

In addition to these two pMSSM points, we study two other representative points: one based on the 
CMSSM and another based on the NUHM. 
The CMSSM is characterized by five parameters: the universal soft mass for scalar particles $m_0$, 
the universal soft gaugino  mass $m_{1/2}$, the universal trilinear coupling $A_0$, the ratio of the two vacuum expectation values 
of the Higgs doublets $\tan\beta = \frac{v_2}{v_1}$ and the sign of the Higgsino mixing mass $\mu$. 
In the so-called NUHM1 model~\cite{nuhm_higgs} there is one extra free parameter, the soft  mass of the Higgs doublets.
In our analysis, this mass is fixed by the absolute value of the Higgsino mixing mass $\mu$  at the electro-weak  symmetry breaking scale,
that we use as input parameter.
The values of the parameters $m_0$, $m_{1/2}$ and $A_0$ for both models are defined at 
the GUT~scale. 

The CMSSM point we are using is defined  as: $m_0=1150$ GeV, $m_{1/2}=1115$ GeV, 
$\tan\beta=40$, $A_0= -2.5\, m_0= -2750$ GeV and sign($\mu$)$>0$.
Using the package {\tt SPheno}~\cite{spheno} this point yields $m_h \simeq 126$ GeV. 
Quite similar results are obtained by using the {\tt FeynHiggs} package~\cite{FH}.
Furthermore, this point is compatible with the WMAP bound on the
neutralino relic density and it delivers acceptable values for the decay $B_s \to \mu^+ \mu^-$ and the
dark matter direct detection cross-section. 
This point belongs  to the stau-coannihilation region of the CMSSM and  has been discussed in~\cite{EO_higgs}.
On the other hand, the NUHM1 point is defined as
$m_0=1000$ GeV, $m_{1/2}=1200$ GeV, $\tan\beta=10$, $A_0= -2.5\, m_0= -2500$ GeV and $\mu=500$ GeV.
At this point  the light Higgs mass is  $m_h=126.5$ GeV and it is also compatible with 
other cosmological and accelerator constraints ~\cite{nuhm_higgs}.
 
\begin{figure}[t!]
\mbox{\hspace{2mm}
\includegraphics[width=7.7 cm]{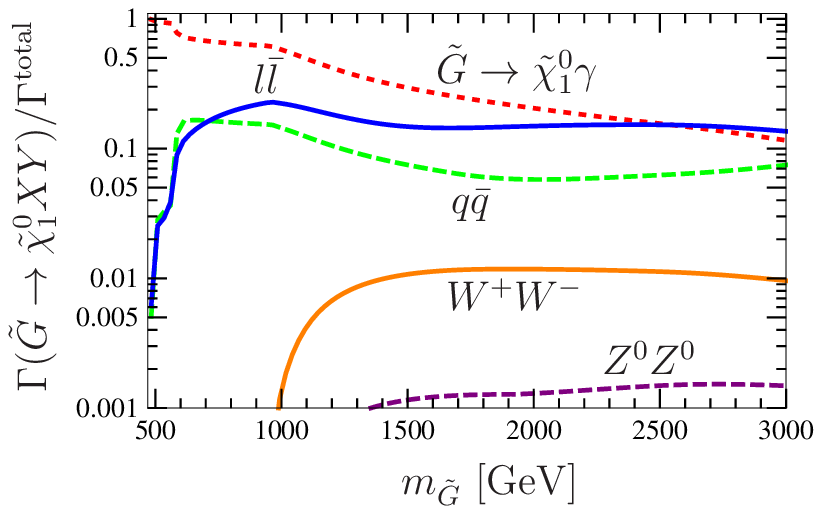}
\hspace{3mm}
\hfill
\includegraphics[width=7.7 cm]{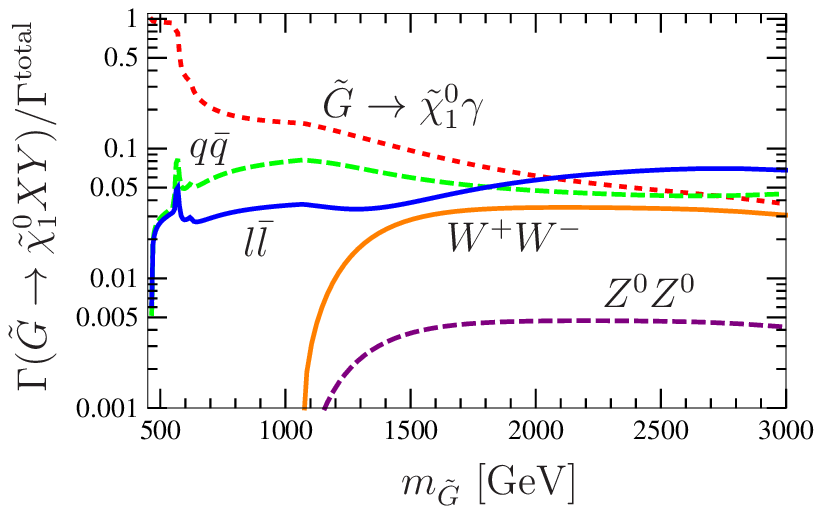}
}
\vspace*{-0.5cm}
\caption[] {The branching ratios for the CMSSM  (left) and NUHM point (right).  The curves are as in Figure~\ref{fig:NeuDec_funnel}  right. }
\label{fig:BR_cmssm_nuhm} 
\end{figure}
In Figure~\ref{fig:BR_cmssm_nuhm} we present the various branching ratios for the important gravitino decay channels, as
functions of the gravitino mass,  as  we have done in Figure~\ref{fig:NeuDec_funnel} right. 
At the CMSSM point the four neutralino masses are $m_{{\tilde{\chi}}^0_{1,2,3,4}}$ $=(485,919,1824,1826) $ ~GeV.
The $\lsp$ is almost a pure bino with its higgsino components less than 0.1\%.
We notice that  the branching ratios for the channels  $\lsp \, l \bar{l} $ and $\lsp \, q \bar{q} $   are approximately of the order of  10\%
after the mass threshold  of the 500 GeV. The two-body channel $\lsp \gamma$ clearly dominates in the
range of small gravitino masses, but after 2000 GeV it becomes comparable to the $\lsp \, l \bar{l} / q \bar{q}$ channels.
On the other hand, the branching ratio for the $\lsp W^+ W^-$ channel is almost an order of magnitude smaller, and  
for the channel $\lsp Z^0 Z^0$ a further order of magnitude smaller. 

Figure~\ref{fig:BR_cmssm_nuhm} right displays the branching ratios for the NUHM point. At this
point $m_{{\tilde{\chi}}^0_{1,2,3,4}}=(470,511,539,1025)$~GeV and there $\lsp$ has a significant higgsino component. 
The two higgsino components ($\tilde H^0_{1,2}$) of the LSP amount to approximately 25\% each.  
In comparison to the CMSSM point, the bino component decreased from almost 100\% to about 50\%. 
The wino component of $\lsp$ is very small.
Therefore, the channels which couple to the bino and wino components, like
$\lsp \, l \bar{l} / q \bar{q}$ except that with stops, are suppressed  by a factor of 2 or more, in comparison to the CMSSM case. 
The channels that depend on the higgsino content of $\lsp$, like  $\lsp W^+ W^-/Z^0 Z^0$ are enhanced by
approximately a factor of 2. For both the CMSSM and NUHM cases displayed in Figure~\ref{fig:BR_cmssm_nuhm}, the
not shown two-body channels, as $\tilde \chi^+_j W^-$,  $\tilde \chi^0_i Z^0$ and $\tilde{g} g $, provide the bulk of the remaining contribution
to the total gravitino decay width.  We notice that also in this case for $\mgrav \gtrsim 2000 $ GeV the  
channels $\lsp \, l \bar{l} / q \bar{q}$  are comparable to the two-body channel $\lsp \gamma$.
  
\begin{figure}[t!]
\begin{center}
\mbox{
\includegraphics[width=8. cm]{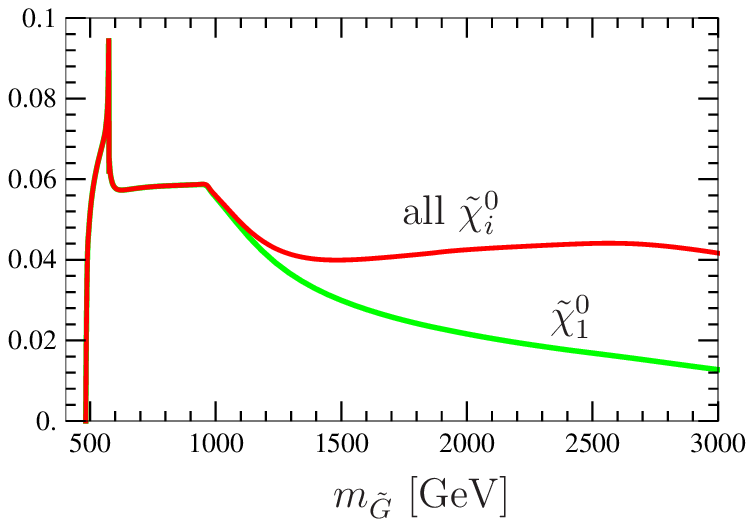}
\hspace{0mm}
\hfill
\includegraphics[width=8. cm]{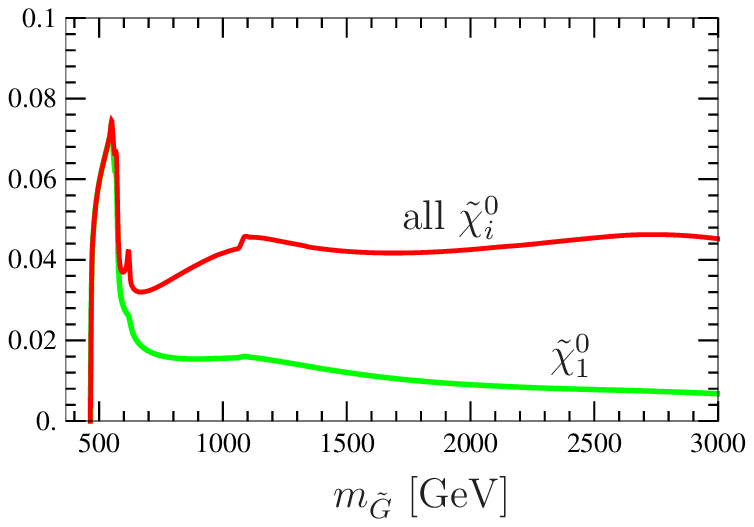}
}
\vspace*{-0.8cm}
\end{center}
\caption[]{The ratio  $\Gamma^{\mathrm{non-reso}} (\grav  \to  \lsp \, X \, Y  ) / \Gamma^{\rm 2B}$  (green curve)  
  and  $\sum_{i=1,..,4}  \Gamma^{\mathrm{non-reso}} (\grav  \to  \tilde \chi_i^0  \, X \, Y  ) / \Gamma^{\rm 2B}$  (red curve),
  for the CMSSM (left) and NUHM (right) point. $\Gamma^{\rm 2B}$ is the total gravitino two-body decay width.}
\label{fig:chi_sum}  
\end{figure}

In Figure~\ref{fig:chi_sum} we present the  ratio  $\Gamma^{\mathrm{non-reso}} (\grav  \to  \lsp \, X \, Y  ) / \Gamma^{\rm 2B}$ (green solid curve)
and the $\sum_{i=1,..,4}  \Gamma^{\mathrm{non-reso}} (\grav  \to  \tilde \chi_i^0  \, X \, Y  ) / \Gamma^{\rm 2B}$  (red solid curve) as function
of the gravitino mass for the CMSSM (left) and NUHM (right) point, we discussed before.
$\Gamma^{\rm 2B}$ denotes the total two-body width of the gravitino.
These figures give us an idea of the contribution of the heavier neutralino states in comparison to $\lsp$, especially for the non-resonant 
contribution of the three-body decay channels $\tilde \chi_i^0  \, X \, Y $. 
The spikes in both cases, just before $\mgrav=600$ GeV are due to the threshold of the two-body decay 
$\grav \to \lsp \, Z^0$. Especially for the NUHM case at right, the second smaller spike corresponds to the almost degenerate 
threshold for the processes $\grav \to \tilde {\chi}_{2,3}^0  \,  Z^0$.
The structures in the region $\mgrav \simeq 1000$ GeV correspond to threshold  $\grav \to \tilde {\chi}_{2}^0 \,  Z^0$,
for the CMSSM case and to  $\grav \to \tilde {\chi}_{4}^0  \, Z^0$ for NUHM.

We see that for $m_{\tilde G}$ at 3 TeV the ratio 
$\sum_{i=1,..,4}  \Gamma^{\mathrm{non-reso}}    (\grav  \to  \tilde \chi_i^0  \, X \, Y  )/ \Gamma^{\rm 2B}$ is four times 
larger than the ratio $\Gamma^{\mathrm{non-reso}} (\grav  \to  \lsp \, X \, Y  )/ \Gamma^{\rm 2B}$, 
both for the CMSSM and the NUHM point. On the other hand, 
for the CMSSM point
for $\mgrav$ up to 1 TeV the ratio $\Gamma^{\mathrm{non-reso}} (\grav  \to  \lsp \, X \, Y  )/ \Gamma^{\rm 2B}$ dominates over the 
other neutralino contributions. For the  NUHM point (right figure)  the dominance of the $\lsp$ ends just before $\mgrav=600$ GeV.
Furthermore, from Figure~\ref{fig:chi_sum}  one can see the size of the total non-resonant contribution 
of the three-body decay  channel $\grav  \to  \tilde \chi_i^0  \, X \, Y  $ relative to the dominant two-body decay width.
For the CMSSM (NUHM)  point, in the region up to  $\mgrav \lesssim 1$~TeV (650 GeV) the non-resonant part of the three-body decays into 
$\lsp$ is about 6\% and decreases to~ 2\%~(1\%) for $\mgrav > 2 $ TeV.

An important comment, however, is in order. The numerical smallness of the contribution of the non-resonant 
three-body decay width can be misinterpreted  to consider the three-body decays to be as unimportant in comparison to the 
two-body decays. Discussing the details of various individual three-body channels in the Figures~\ref{fig:ww}--\ref{fig:Zh0} we 
have seen that below the two-body mass thresholds, the exact knowledge of the gravitino  three-body decay widths  is important 
in the calculation of the corresponding branching ratios in the whole range of $\mgrav$. Thus,
it is essential to calculate {\em all} possible  three-body decay channels for the process $\grav  \to  \tilde \chi_i^0  \, X \, Y  $, in order
to have the complete information both for the total gravitino width and the individual branching ratios,  for any gravitino mass. 

\begin{figure}[t!]
\begin{center}
\mbox{\hspace{2mm}
\includegraphics[width=7.7 cm]{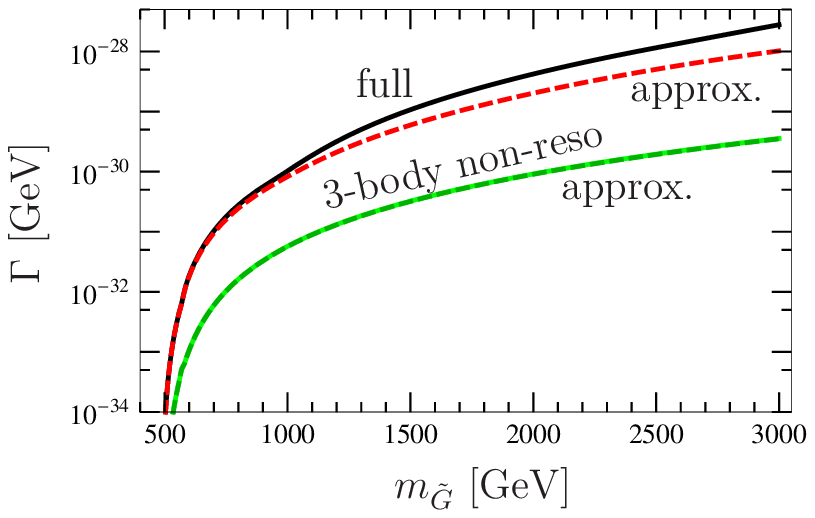}
\hspace{3mm}
\hfill
\includegraphics[width=7.7 cm]{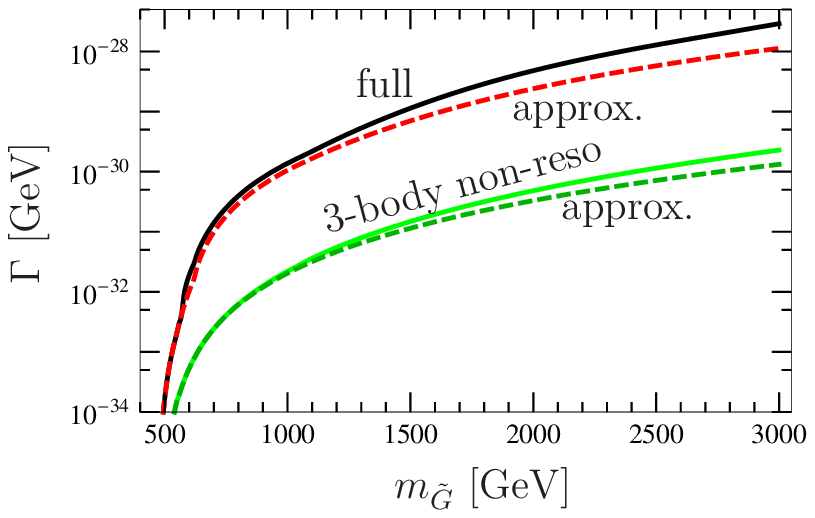}
}

\vspace*{-0.5cm}
\end{center}
\caption[]{Comparison between the full gravitino width, that is the contribution from the 
 two-body decays and the three-body  non-resonant part (black curve) and the approximation (red dashed curve).
 The  green curve is the three-body non-resonant part only and the dark green  dashed curve the corresponding approximation. 
 The left  (right) figure correspond to the CMSSM (NUHM) point as described in the text.  }
\label{fig:app}  
\end{figure}

In Figure~\ref{fig:app} we compare the full result for the gravitino decay width (black curve), that corresponds
to the sum of the two-body decays and the non-resonant part of the three-body decays   $ \grav  \to  \tilde \chi_i^0  \, X \, Y  $,
with a commonly used approximation (red dashed curve). In this approximation  one calculates 
the gravitino width using the two-body channels $\tilde \chi_i^0 (\gamma, \, Z^0,  \,  h^0,  \,  H^0,  \, A^0) $ and the 
three-body channels $\lsp \, f \bar{f}$, where $f$ can be either lepton or quark. 
The approximation is relatively good up to $\mgrav =1000$ GeV, differing from the full result by $\sim 20\%$, for the CMSSM point. 
For the NUHM point this difference is $\sim 25\%$. But for larger gravitino masses ($\mgrav \simeq 3000$ GeV),
where almost all two-body channels contribute the difference is quite important. The approximation underestimates 
the full result by a factor of 2.7 for the CMSSM point and by 2.9 for the NUHM. 
Comparing the non-resonant contribution of the three-body decays we get a different picture. For
the CMSSM point,  the approximation (dark green dashed curve) describe quite well the full non-resonant part (light green solid curve).
The reason for this is that the dominant bino content of $\lsp$ leads to the dominance of the three-body decay channel $\lsp \, f \bar{f}$,
as we have seen also in Figure~\ref{fig:BR_cmssm_nuhm}.
On the other hand, at the NUHM point the approximation are equally important, the approximation is  almost a factor of 2 
smaller than the complete result for $\mgrav=3$ TeV.
This is due to the fact that the 
three-body channels  $\lsp \, W^+\, W^-/Z^0\, Z^0$ etc. are not included in the approximation but play an important role 
in this scenario.   

\begin{figure}[t!]
\begin{center}
\includegraphics[width=8.cm]{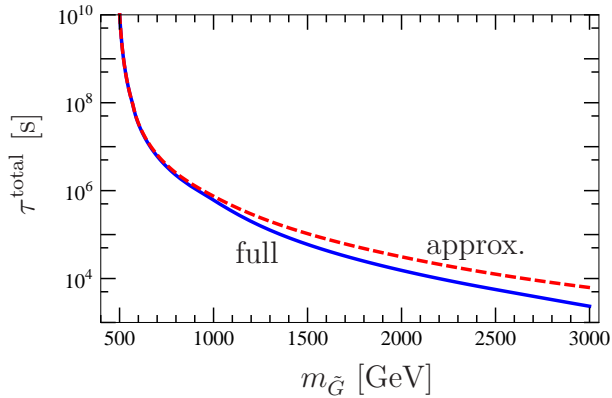}
\end{center}
\vspace*{-0.5cm}
\caption[]{The gravitino lifetime as a function of the gravitino mass for the CMSSM point.
The blue curve is the lifetime as calculated including all the two-body and three-body channels in our study
and the red dashed curve  corresponds to the approximation described in the text to Figure~\ref{fig:app}.  }
\label{fig:life}  
\end{figure}

The relevance of the calculation of the full three-body gravitino decay width is also underlined
in Figure~\ref{fig:life}. For the CMSSM point, we display the gravitino lifetime in seconds using the full 
result (blue curve) and the approximation described above (red dashed curve). The corresponding figure for the NUHM point is quite similar.
Reflecting the behavior of  Figure~\ref{fig:app} the approximation up to $\mgrav=1$ TeV is relatively good, but for larger gravitino masses
overestimates the gravitino lifetime by a factor of 2 and more.
As already mentioned, the precise calculation of the  gravitino lifetime especially in the range $\mgrav\sim 2-4 $ TeV,
is important because one thinks that the models with unstable gravitinos can explain the so-called Lithium cosmological puzzle~\cite{Cyburt:2009pg}.
This is due to the fact that the lifetime of the gravitino for these gravitino masses is $\sim 10^4$~s and shorter, time scales that can be 
quite relevant to BBN predictions for the primordial light elements abundances, and especially for the \li6 and \li7 isotopes. 
Thus, our calculation is of importance in the phenomenology of supersymmetric models with unstable gravitinos. 

\begin{figure}[t!]
\begin{center}
\includegraphics[width=8.cm]{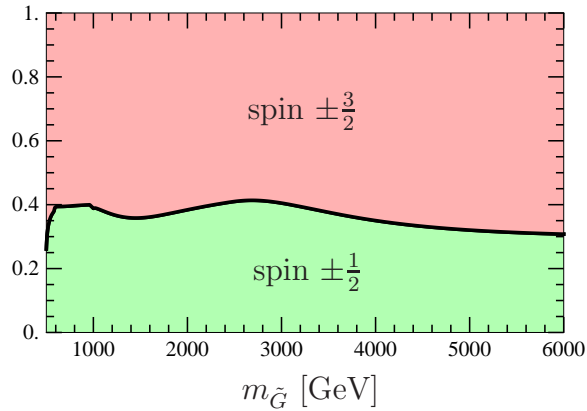}
\end{center}
\vspace*{-0.5cm}
\caption[]{The relative contribution of the $\pm \frac{1}{2}$   gravitino states (green region up to the black curve) to the total gravitino width,
as a function of the gravitino mass. 
The remaining part (salmon
color) corresponds to the relative contribution of the 
$\pm \frac{3}{2}$  
 spin states to the total gravitino decay width. For this figure the CMSSM point parameters are used. }
\label{fig:hel}  
\end{figure}

The use of the Weyl-van-der-Waerden formalism for the calculation of the gravitino width
allows one to explicitly compute the individual contributions of the four gravitino spin states to the total width.
It is known that the gravitino in broken superysymmetric models as studied here, 
acquires its mass by absorbing the goldstino $\pm \frac{1}{2}$ modes.

In  Figure~\ref{fig:hel}, we display
the sum of the contributions of the $\pm \frac{1}{2}$ gravitino spin states (green region), 
relatively to the sum of the $\pm \frac{3}{2}$   contributions (salmon region) for the CMSSM point. 
Again the NUHM case is very similar and thus it is not shown. 
We see that for $\mgrav$ up to about 3~TeV the  $\pm \frac{1}{2}$ states contribute up to 40\% to the total decay width. The remaining 60\% 
is due to the  $\pm \frac{3}{2}$ spin states. Eventually, for larger gravitino masses the $\pm \frac{1}{2}$ spin states contribute about 30\% of 
total width, this contribution remaining constant for much larger values of the gravitino mass. 
The reason for this is that in the gravitino decay amplitudes the center of mass energy $\sqrt{s}$ of the 
process is actually the gravitino mass $\mgrav$.
Hence, we are far from the high-energy limit ($\sqrt{s} \gg \mgrav$),  
in which one expects the goldstino ($\pm \frac{1}{2}$) components to be dominant.

\section{Conclusions }
\label{sec:conclusions}

We have presented results from a new calculation of the gravitino decay width in the context of supersymmetric models, where
$R$-parity is conserved. This calculation includes all two-body decay channels 
$\grav \to \widetilde{X}\, Y$ where $\widetilde{X}$ is a sparticle decaying as 
$\widetilde{X} \to \lsp \, X$. $X$ and $Y$ are SM particles,  and $\lsp$ is the 
lightest neutralino that plays the role of the LSP. In addition, we have also calculated
the contributions of the three-body decay channels
$\grav \to \tilde \chi^0_i \, X \, Y$. Especially  the computation of the three-body decays is quite a complicated task.
For this purpose we have used the packages {\tt FeynArts} and {\tt FormCalc} that perform symbolic analytic calculations.
In those packages we have added appropriate modules in order to accommodate them for particles with spin 3/2 like the gravitino.
Furthermore, we have used the Weyl-van-der-Waerden formalism which considerably simplifies the treatment of 
the complex spin structure of the three-body gravitino decay amplitudes and enables us to calculate the 
individual contributions of each gravitino spin state to the total decay width. 
 
An important advantage of our calculation of the gravitino  three-body decay amplitudes, is 
that it  treats automatically the intermediate exchanged particles  that are on their mass shell by using the narrow width approximation.
In such a way, we not only avoid the double counting among the two- and three-body channels,  
but we also improve the performance of the numerical phase space integration over the unstable exchanged particles in the propagators.
  
In our numerical analysis, we have chosen four representative points: two points based on the pMSSM model and 
two other based on the CMSSM and NUHM1 models. These points satisfy all the recent high energy experimental constraints 
from LHC and LHCb, but also the cosmological constraints from WMAP, Planck, and XENON100.  

We have found that the knowledge of the complete three-body decay amplitude with a neutralino is important in order 
to compute precisely the gravitino width for any gravitino mass, not only below the mass thresholds of the   
dominant two-body decay channels, but also above them. Moreover, comparing the full 
gravitino decay width, up to the three-body level, with a previously used approximation we have demonstrated 
that the full result can be quite different, especially for $\mgrav$ in the range 2--4 TeV.  This difference is reflected
also in the gravitino lifetime that we know is an important parameter to constraint models with unstable gravitinos adopting
the BBN predictions. 
In addition, we have found that all the gravitino spin states ($\pm {1 \over 2}$ and $\pm {3 \over 2}$) contribute at the same order
of magnitude, even for quite large gravitino masses.
 
Our results have been calculated using the code {\tt GravitinoPack} that will be publicly released soon. This package
incorporates various {\sc Fortran} and {\sc Mathematica} routines that compute all the discussed two- and three-body gravitino branching 
ratios and the gravitino decay width. In addition, {\tt GravitinoPack} will calculate 
analogous decay widths for the complementary case where the gravitino is stable and the LSP. 

\vspace*{1cm}

\section*{Acknowledgements}

This work is supported by the ÓFonds zur F\"orderung der wissenschaftlichen Forschung (FWF)Ó of Austria, 
project No. I 297-N16. The authors thank Thomas Hahn for helpful discussions regarding the implementation of particles with spin 3/2
into FeynArts and they are grateful to Walter Majerotto for the careful reading of the manuscript.
V.C.S. was supported by Marie Curie International Reintegration grant SUSYDM-PHEN,
MIRG-CT-2007-203189.




\end{document}